\documentclass[fleqn,usenatbib]{mnras}

\usepackage[T1]{fontenc}
\usepackage{ae,aecompl}

\usepackage{graphicx}	
\usepackage{amsmath}	
\usepackage{amssymb}	
\usepackage{gensymb} 
\usepackage{newtxtext,newtxmath}
\usepackage{hyperref}   
\usepackage{subcaption}
\usepackage{tabularx}
\usepackage{xcolor}

\title[VST ATLAS Quasar Survey I]{The VST ATLAS Quasar Survey I: Catalogue}

\author[A. M. Eltvedt et al.]{
Alice M. Eltvedt$^{1}$\thanks{E-mail: alice.m.eltvedt@durham.ac.uk},
T. Shanks$^{1}$,
N. Metcalfe$^{1}$,
B. Ansarinejad$^{1,2}$,
L. F. Barrientos$^{3,4}$,
R. Sharp$^{5}$, \newauthor
U. Malik$^{5}$,
D.N.A. Murphy$^{4}$,
M. Irwin$^{4}$,
M. Wilson$^{1}$,
D. M. Alexander$^{1}$,
Andras Kovacs$^{6,7}$,
Juan Garcia-Bellido$^{8}$, \newauthor
Steven Ahlen$^{9}$,
David Brooks$^{10}$,
Axel de la Macorra$^{11}$,
Andreu Font-Ribera$^{12}$,
Satya Gontcho a Gontcho$^{13,14}$, \newauthor
Klaus Honscheid$^{15,16}$,
Aaron Meisner$^{17}$,
Ramon Miquel$^{18,19}$,
Jundan Nie$^{20}$,
Gregory Tarlé$^{21}$, \newauthor
Mariana Vargas-Magaña$^{11}$, 
Zhimin Zhou$^{20}$
\\
$^{1}$Centre for Extragalactic Astronomy, Department of Physics, Durham University, South Road, Durham, DH1 3LE, UK\\
$^{2}$ School of Physics, University of Melbourne, Parkville, VIC 3010, Australia\\
$^{3}$Instituto de Astrofisica, Facultad de Fisica, Pontificia Universidad Catolica de Chile, Santiago, Chile\\
$^{4}$Institute of Astronomy, University of Cambridge, Madingley Road, Cambridge CB3 0HA, UK \\
$^{5}$Research School of Astronomy and Astrophysics, Australian National University, Canberra, ACT 2611, Australia\\
$^{6}$Departamento de Astrof´ısica, Universidad de La Laguna (ULL), E-38206, La Laguna, Tenerife, Spain\\
$^{7}$Instituto de Astrof´ısica de Canarias, C/ V´ıa L´actea, s/n, 38205 San Crist´obal de La Laguna, Santa Cruz de Tenerife, Spain\\
$^{8}$Instituto de Física Teórica UAM/CSIC, Universidad Autónoma de Madrid, 28049 Madrid, Spain\\
$^{9}$Physics Dept., Boston University, 590 Commonwealth Avenue, Boston, MA 02215, USA\\
$^{10}$Department of Physics \& Astronomy, University College London, Gower Street, London, WC1E 6BT, UK\\
$^{11}$Instituto de Física, Universidad Nacional Autonoma de México, Cd. de México C.P. 04510, México\\
$^{12}$Institut de F´ısica d’Altes Energies (IFAE), The Barcelona Institute of Science and Technology, Campus UAB, 08193 Bellaterra
Barcelona, Spain\\
$^{13}$Lawrence Berkeley National Laboratory, One Cyclotron Road, Berkeley, CA 94720, USA\\
$^{14}$Department of Physics and Astronomy, University of Rochester, 500 Joseph C. Wilson Boulevard, Rochester, NY 14627, USA\\
$^{15}$Center for Cosmology and Astro-Particle Physics, The Ohio State University, Columbus, OH 43210, USA\\
$^{16}$Department of Physics, The Ohio State University, Columbus, OH 43210,
USA\\
$^{17}$NSF’s National Optical-Infrared Astronomy Research Laboratory, 950 N. Cherry Avenue, Tucson, AZ 85719, USA\\
$^{18}$Instituci´o Catalana de Recerca i Estudis Avan¸cats, Passeig de Llu´ıs Companys, 23, 08010 Barcelona, Spain\\
$^{19}$Institut de F´ısica d’Altes Energies (IFAE), The Barcelona Institute of Science and Technology, Campus UAB, 08193 Bellaterra
Barcelona, Spain\\
$^{20}$National Astronomical Observatories, Chinese Academy of Sciences, A20 Datun Rd., Chaoyang District, Beijing, 100012, P.R. China\\
$^{21}$Department of Physics, University of Michigan, Ann Arbor, MI 48109, USA\\
}

\date{Accepted XXX. Received YYY; in original form ZZZ}

\pubyear{2022}

\begin{document}
\label{firstpage}
\pagerange{\pageref{firstpage}--\pageref{lastpage}}
\maketitle

\begin{abstract}
We present the VST ATLAS Quasar Survey, consisting of $\sim1,229,000$ quasar (QSO) candidates with $16<g<22.5$ over $\sim4700$ deg$^2$. The catalogue is based on VST ATLAS$+$NEOWISE imaging surveys and aims to reach a QSO sky density of $130$ deg$^{-2}$ for $z<2.2$  and $\sim30$ deg$^{-2}$ for $z>2.2$. One of the aims of this catalogue is to select QSO targets for the 4MOST Cosmology Redshift Survey. To guide our selection, we use  X-ray/UV/optical/MIR data in the extended William Herschel Deep Field (WHDF) where we find a $g<22.5$ broad-line QSO density of $269\pm67$ deg$^{-2}$, roughly consistent with the expected $\sim196$ deg$^{-2}$.  We also find that $\sim25$\% of our QSOs are morphologically classed as optically extended. Overall, we find that in these deep data, MIR, UV and X-ray selections are all $\sim70-90$\% complete while X-ray suffers less contamination than MIR and UV. MIR is however more sensitive than X-ray or UV to $z>2.2$ QSOs at $g<22.5$ and the  $S_X(0.5-10\,{\rm keV})>1\times10^{-14}$ ergs cm$^{-2}$ s$^{-1}$ limit of eROSITA. We then adjust the  selection criteria from our previous 2QDES pilot survey and prioritise VST ATLAS candidates that show both UV and MIR excess, while also selecting candidates initially classified as extended. We test our selections using data from DESI (which will be released in DR1) and 2dF  to estimate the efficiency and completeness of our selections, and finally we use ANNz2 to determine photometric redshifts for the QSO candidate catalogue. Applying over the $\sim4700$ deg$^2$ ATLAS area gives us $\sim917,000$  $z<2.2$ QSO candidates of which 472,000 are likely to be $z<2.2$ QSOs, implying a sky density of $\sim$100 deg$^{-2}$, which our WHDF analysis suggests will rise to at least 130 deg$^{-2}$ when eROSITA X-ray candidates are included. At $z>2.2$, we find $\sim310,000$ candidates, of which 169,000 are likely to be QSOs for a sky density of $\sim36$ deg$^{-2}$. 

\end{abstract}




\section{Introduction}
\label{sec:Introduction}

QSOs are the most luminous subset of Active Galactic Nuclei (AGN), which are powered by accretion onto a black hole. Here we develop selection criteria for a photometrically selected QSO catalogue based on VST ATLAS \citep{Shanks2015} $+$unWISE neo6 \citep{unWISE2019}. We aim to achieve a sky density at  $g<22.5$ of 130 deg$^{-2}$ at $z<2.2$ and 30 deg$^{-2}$ at $z>2.2$ over $\approx4700$ deg$^2$, comparable to the sky densities projected by the Dark Energy Spectroscopic Instrument experiment (DESI) \citep{DESI2016} and observationally confirmed by \cite{Chaussidon2022}. We utilize methods outlined in \citet{Chehade2016} and develop further selection techniques by comparing our results to  X-ray QSOs from \citet{Bielby2012} in the WHDF \citep{Metcalfe2001}, and preliminary DESI data from DESI DR1.

This catalogue aims to be part of the spectroscopic fiber targeting of the upcoming 4MOST Cosmology Redshift Surveys \citep{4MOST2019}, where it will be combined with 2800 deg$^{2}$ from the Dark Energy Survey (DES) \citep{DES2016} to give 7500 deg$^2$ at 130 deg$^{-2}$ for QSO cosmology projects. It could also be used to target eROSITA AGN surveys \citep{eROSITA2012}. The eROSITA X-ray AGN survey has average resolution of only $\approx20''$ so our optical/MIR catalogue will also help target fibres for spectroscopic follow-up with 4MOST in our overlap areas. The long-term aim of this QSO survey is to probe the nature of dark energy and dark matter by primarily comparing gravitational lensing and redshift space distortion analyses (e.g. \citealt{Kaiser1987}) but also via BAO using QSOs as tracers at $z<2.2$ and the Lyman-$\alpha$ forest at $z>2.2$ from the final 4MOST redshift surveys. The dark energy equation of state will thus be measured and  tests of modified gravity models as an alternative explanation of the accelerating Universe will also be made. In the VST ATLAS QSO Survey paper II we shall report on the lensing of VST ATLAS QSOs by foreground galaxies and galaxy clusters. We also detect lensing of the cosmic microwave background (CMB) by the QSOs. 

The outline of this paper is as follows. In Section \ref{sec:data} we describe the imaging and spectroscopic surveys we use to create and test our QSO catalogue. We describe QSO selection methods based on the 2QDESp and WHDF surveys, which we utilize to start, test, and adapt our QSO selections in Section \ref{sec:qso_selection_ini}. Section \ref{sec:final_selections} details the final VST-ATLAS QSO catalogue selections. Section \ref{sec:spectroscopic_completeness} contains a spectroscopic completeness analysis of our VST-ATLAS QSO catalogue, using preliminary DESI data as well as our own preliminary data from AAT 2dF. We present the final VST-ATLAS QSO candidate catalogue in Section~\ref{sec:final_catalogue}. Finally, we finish our analysis in Section \ref{sec:ANNZ} where we utilize the ANNz2 photometric redshift code to determine a $z<2.2$ and $z>2.2$ redshift sample. We discuss our results in Section \ref{sec:conclusions}.

\section{Data}
\label{sec:data}

\subsection{Imaging Surveys}

\subsubsection{VST-ATLAS}
\label{sec:ATLAS} 

The ESO VST ATLAS data we utilize in this work is from the DR4 ATLAS catalogue released in 2019. ATLAS is a photometric survey which images $\sim4700$ deg$^{2}$ of the Southern sky ($\approx$2000 deg$^2$ in the Northern Galactic Cap, NGC, and $\approx$2700 deg$^2$ in the Southern Galactic Cap, SGC, in the $ugriz$ bands, designed to probe similar depths as the Sloan Digital Sky Survey (SDSS) \citep[e.g.][]{SDSS2000}. The imaging was performed with the VLT Survey Telescope (VST), which is a 2.6-m wide-field survey telescope with a $1^{\circ}\times1^{\circ}$ field of view. It is equipped with the OmegaCAM camera (\citealt{Kuijken2002}), which is an arrangement of 32 CCDs with $2k\times4k$ pixels, resulting in a $16k\times16k$ image with a pixel scale of $0.''21$. The two sub-exposures taken per 1 degree field are processed and stacked by the Cambridge Astronomy Survey Unit (CASU). This pipeline provides catalogues with approximately $5\sigma$ source detection that include fixed aperture fluxes and morphological classifications. The processing pipeline and resulting data products are described in detail by \cite{Shanks2015}. We create band-merged catalogues using {\sc TOPCAT} \citep{TopCat2005}. For our quasar catalogue, we utilize a $1.''0$ radius aperture (aper3 in the CASU nomenclature) as well as the Kron magnitude in the $g-$band, and the morphological star-galaxy classification supplied as a default in the CASU catalogues in the $g-$band. This classification is discussed in detail by \cite{Gonzalez2008}. The $u$-band data in DR4 consist of $2\times120$s exposures in the $\approx700$ deg$^{-2}$ area at  Dec$<-20$ deg in the NGC and $2\times60$s exposures elsewhere. We utilize the $2\times60$s $u$-band exposures of the complementary ATLAS Chilean Survey (ACE, Barrientos et al in prep.) to increase the $u$-band exposure time to 240s exposure throughout the entire DR4 area. We combine the ATLAS and Chilean $u-$band data by averaging their magnitude values weighted by the relative seeing on the two exposures. Approximately $1000$ deg$^2$ of the DR4 SGC area and NGC area at Dec>-20 deg did not have Chile $u$-band data at the time of this work. In these areas we simply use the shallower ATLAS DR4 data. To ensure as many objects as possible have $u$-band measurements, we do not detect objects independently on the $u$ images but instead we 'force' photometry at the positions of all the $g$-band detections. To avoid problems with detector saturation at brighter magnitudes, in what follows we restrict the ATLAS data to objects with $g>16$. The area covered by VST ATLAS, as well as the surveys we are utilizing in the analyses of this paper can be seen in Fig.~\ref{fig:DESI_ATLAS_area}.

\subsubsection{NEOWISE}
\label{Sec:unWise_neo6}
The NASA satellite Wide-field Infrared Survey Explorer (WISE) \citep{Wright2010}, mapped the entire sky in four pass-bands $W1$, $W2$, $W3$, and $W4$ at $3.4$, $4.6$, $12$, and $22 \mu m$ respectively, with $5\sigma$ point source limits at $W1$ = 16.83 and $W2$ = 15.60 mag in the Vega system.
The unWISE catalogue \citep{unWISE2019} presents $\sim$2 billion objects observed by WISE, with deeper imaging and improved modelling over AllWISE, 
detecting sources approximately $0.7$ magnitudes fainter than AllWISE in $W1$ and $W2$, ie $5\sigma$ limits of $W1=17.5$ and $W2=16.3$ in the Vega system. This deeper imaging is made possible through the coaddition of all available $3 - 5 \mu m$ WISE imaging, including that from the ongoing NEOWISE-Reactivation mission, increasing the total exposure time by a factor of $\sim5$ relative to AllWISE \citep{unWISE2019}. We use the pre-release version of DR3 of the unWISE catalogue (neo6), provided by E. Schlafly, in this work. 


To allow checks of unWISE quasar selection, we also download data from the DECaLS Legacy Survey DR9 release \citep{DECaLS2019} as this is the data which has been used by \cite{DESI2016} in their science, targeting, and survey design. 
This includes the $W1$ and $W2$ WISE fluxes using `forced' photometry at the locations of Legacy Surveys optical sources in the unWISE maps. Being `forced', these data go somewhat deeper than the unWISE neo6 catalogue, but, of course, only exist for objects with optical photometry.

\subsubsection{William Herschel Deep Field (WHDF)}
\label{sec:WHDF_intro}

To perform an analysis of X-ray selected quasars, we use the William Herschel Deep Field (WHDF) data provided by \cite{Metcalfe2001}. This data covers a $16'\times16'$ area of sky with data in the $UBRIZHK$ bands and goes several magnitudes deeper than our VST ATLAS data. Unfortunately, particularly for $U$ and $B$, the passbands are very different from those used in the VST ATLAS survey. To overcome this we matched to the SDSS Stripe 82 photometry (described in \citealt{Pier_2003}), whose passbands are very similar to VST ATLAS. Although this is less deep than the WHDF photometry, for $B<23.5$, $\approx$95\% of our WHDF objects have Stripe 82 photometry. We retain the star/galaxy separation information from the deeper, original, WHDF data.

This is then combined with a 75ksec Chandra ACIS-I X-ray exposure \citep{Vallbe2004,Bielby2012} and the mid-infrared (MIR) 3.6 and $4.5\mu m$ Spitzer SpIES data  \citep{Timlin2016} to provide 0.5-10 keV X-ray fluxes and the equivalent of $W1$ and $W2$ band magnitudes.



\subsection{Spectroscopic Surveys}

\subsubsection{2QZ}

The 2dF QSO Redshift Survey (2QZ; \citep{Boyle2002,Croom2005}, covers approximately 750 deg$^{2}$ of the sky, with $\approx480$ deg$^2$ overlap with VST-ATLAS. It used the 2-degree Field (2dF) multi-object spectrograph at the Anglo Australian Telescope (AAT) to target sources, and discovered $\approx23000$ QSOs at $z<3$. The areas targeted for 2QZ are contained within the 2dF Galaxy Redshift Survey sky coverage (\cite{Colles2001}, 2dFGRS). The 2QZ catalogue utilizes photometric colour cuts to select QSO targets. Therefore, we can use the 2QZ quasar catalogue to test for completeness of our new catalogue as it spans a redshift range of $0.3<z<2.2$, which includes our target redshift range. At higher redshifts, the completeness of the 2QZ survey rapidly drops as the Lyman-alpha forest enters the u-band. Additional incompleteness may be due to AGN dust absorption. See \cite{Croom2005} for further description of the 2QZ QSO survey.

\subsubsection{2QDESp}
 
The 2QDES Pilot Survey (2QDESp) \citep{Chehade2016} was the first survey to use VST ATLAS photometry to target QSOs. They attempted to target QSOs up to $g<22.5$, with high completeness up to $g\approx20.5$ with an average QSO sky density of $\approx$70 deg$^{-2}$ in the redshift range of $0.8<z<2.5$. The target depth of $g\leq 22.5$ was to probe the clustering properties of intrinsically faint quasars as this was a relatively unexplored depth for the targeted redshift range at that time. 
\newline 

As 2QDESp used VST ATLAS data, albeit an earlier release, we will base our selection methods on the 2QDESp selection criteria as we aim to find sources at these faint magnitudes with a higher sky density. We are able to select fainter targets as we use the unWISE catalogue in conjunction with VST ATLAS photometry, instead of the AllWISE all-sky source catalogue used by 2QDESp. We also have deeper $u$-band data and the DR4 release encompasses the full ATLAS area, which was not completed at the time of 2QDESp.

\subsubsection{DESI}

The Dark Energy Spectroscopic Instrument (DESI) \citep{DESI2016} is a Stage IV dark energy measurement using baryon acoustic oscillations and other techniques that rely on spectroscopic measurements. The main spectroscopic survey is conducted on the Mayall 4-metre telescope at Kitt Peak National Observatory. Based on DECaLS DR9 photometry, DESI has a target depth of $r<23$. We utilize main survey data in the seventh internal data release of DESI spectra, Guadalupe (which will be released in DR1), to check our VST-ATLAS photometry as well as our QSO candidate selection via DESI spectroscopy. This data has an $\approx$144 deg$^2$ overlap with ATLAS.  We shall also use DESI  Guadalupe spectroscopy covering the WHDF to increase the numbers of known quasars with redshifts in the WHDF area, beyond those previously reported by \cite{Vallbe2004} and \cite{Bielby2012}.

\subsubsection{2dF}

We are able to test our final QSO selection using the 2-degree Field (2dF) fibre coupler feeding 392 fibres over 3 deg$^2$ into the AAOmega  spectrograph \citep{Sharp2006} at the Anglo Australian Telescope (AAT). The spectrograph uses a dichroic beam-splitter at 5700\AA~ and the fibres have a $2.''1$ diameter. We utilize the multi-object mode and the 580V and 385R gratings, giving a wavelength range from 3700-8800\AA~ with a spectral resolution of $\approx$1300. We observed two trial VST ATLAS fields, NGC-F1 and NGC-F2, with observational details are given in Section  \ref{sec:2dF_comparison}.

\begin{figure*}
	\centering
	\includegraphics[width=\textwidth]{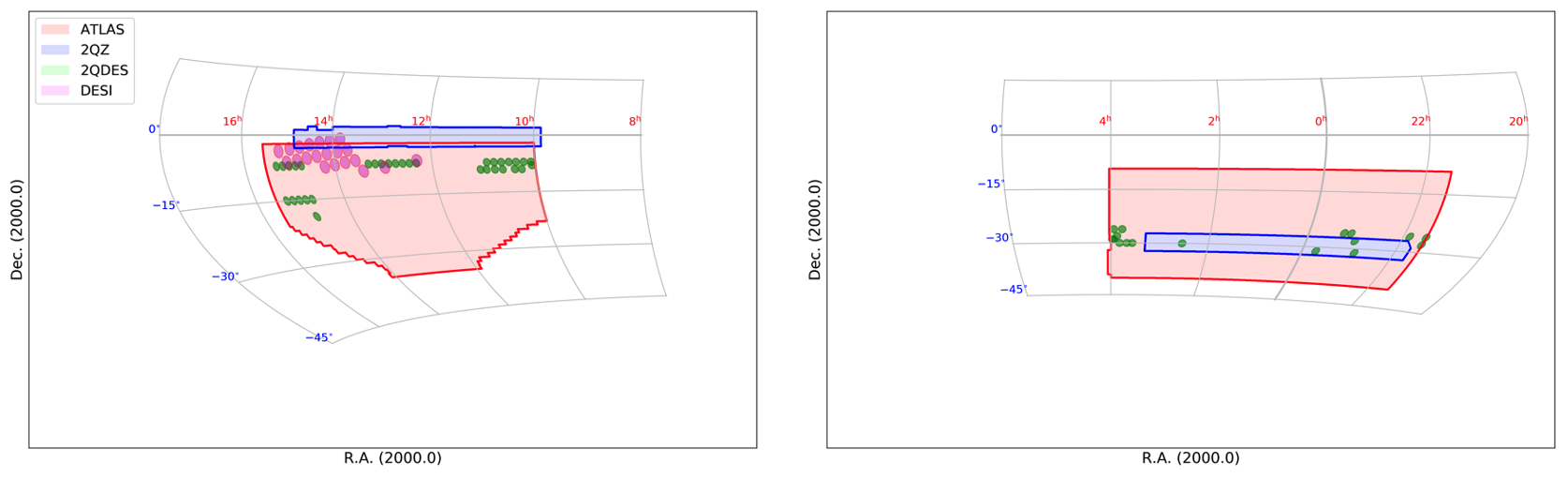}
	\caption[survey overlap]{The sky coverage of VST ATLAS NGC is shown in red, including 2000 deg$^2$ in the NGC in the left hand panel and 2700 deg$^2$ in the SGC in the right hand panel. The map also shows areas where other surveys used in this work overlap  VST ATLAS. The 2dF Quasar Survey (2QZ, \citealt{Croom2005}) area is shown in blue, the 2dF QSO Dark Energy Survey pilot (2QDESp,     \citealt{Chehade2016}) area in green and  the area covered by DESI by the time of the internal release used here, in magenta.}
	\label{fig:DESI_ATLAS_area}
\end{figure*}

\section{Optimizing QSO Selection via 2QDES + WHDF}
\label{sec:qso_selection_ini}

To create the VST ATLAS QSO catalogue, we start from photometric selection methods in multiple colour spaces based on previous work using VST ATLAS$+$AllWISE catalogues. We utilize both the ultra-violet excess (UVX) and the mid-infrared excess (MIRX) properties of QSOs to create photometric colour cuts for our target selection, following \cite{Chehade2016}. We test the completeness of these selections using QSO idendified in the deeper WHDF data at X-ray, optical, and MIR wavelengths. We then adjust these improved selections to allow for the brighter flux limits that apply to VST ATLAS and unWISE relative to the WHDF, always aiming to minimise stellar and galaxy contamination while maximizing completeness of the quasar sample. We perform these colour cuts in the regions covered by VST-ATLAS and unWISE in both the NGC and SGC survey areas in the Southern hemisphere. UVX colour cuts were previously used by 2QZ and SDSS \citep{SDSSQSO} to select quasars in the redshift range of $z<2.2$. We then follow \cite{Chehade2016} in combining UV and MIR photometry to make our QSO selections. The continued inclusion of  UV criteria differentiates this work from e.g. the extended Baryon Oscillation Spectroscopic Survey (eBOSS) \citep{eBOSS2016} and DESI who use only MIRX selection \citep{Yeche2020}. We shall use  spectroscopic surveys such as 2QZ, 2QDESp, eBOSS, DESI and new 2dF observations to optimise the ATLAS selection and compare selection efficiencies with results from these other spectroscopic surveys.

\subsection{2QDESp QSO Selection}
\label{sec:2QDESp_cuts}

Our initial ATLAS selections are based on the UVX and MIRX QSO selections made by \cite{Chehade2016} for 2QDESp, with the deeper NEOWISE (neo6) replacing AllWISE as the MIR survey. Their UVX/optical selections were made in the $u-g:g-r$ and $g-r:r-i$ colour spaces and their MIRX-optical selections are made in the $g-i:i-W1$ and $g:W1-W2$ colour spaces. \cite{Chehade2016} utilize a combination of VST-ATLAS and WISE photometry in $\approx$150 deg$^2$ of the Southern hemisphere for their analysis, complementing their selection with the XDQSO code for quasar classification. We expand and improve this photometric selection by using VST-ATLAS over $\approx$4700 deg$^2$ of the Southern hemisphere. This paper does not include an XDQSO selection as it would require some re-calibration for the deeper photometry in the $u-$, $W1-$, and $W2-$ bands we are utilizing. 

The original colour selections from \cite{Chehade2016} are as follows. The VST ATLAS photometry is in AB magnitudes and the unWISE photometry is in Vega magnitudes. The UVX/optical selection is given in Eq. \ref{eq_chehade_ugr}:

\vspace{-0.3cm}
\begin{equation}
\begin{aligned}
& -1.0\leq (u-g) \leq 0.8 \,\, \\
& -1.25\leq (g-r) \leq 1.25  \,\, \\
& (r-i)\geq 0.38-(g-r) 
\end{aligned}
\label{eq_chehade_ugr}
\end{equation}

\noindent The selections exploiting mid-IR excess are given in Eq. \ref{eq_chehade_giw1}:

\vspace{-0.3cm}
\begin{equation}
\begin{aligned}  
&  (i-W1)\geq (g-i)+1.5\,\,\&-1.0\leq (g-i)\leq1.3\,\,\&\,\,(i-W1) \leq8\,\\
&  \&\, [(W1- W2)>0.4\,\,\&\,\,(g<19.5)\,\, || \\
&  (W1- W2)>-0.4*g+8.2 \,\,\&\,\,  (g>19.5)] \\
\end{aligned}
\label{eq_chehade_giw1}
\end{equation}

These selections are graphically displayed in Fig. 1 of \cite{Chehade2016}. Following the colour selections outlined above, we note that the maximum confirmed quasar sky density achieved by \cite{Chehade2016} was $\sim$90 deg$^{-2}$ for $z<2.2$ QSOs. This leaves us below our target density of 130 deg$^{-2}$ at $z<2.2$ (plus 30 deg$^{-2}$ at $z>2.2$), motivating us to further improve these selections and use them in conjunction with better data.

\subsection{William Herschel Deep Field (WHDF) QSO Selection}

Our first attempt to refine our QSO selection is based on objects in the extended William Herschel Deep Field (\citealt{Metcalfe2001, Metcalfe2006}). Although a small, $\sim16'\times16'$ area, here we have high signal-to-noise optical data which is several magnitudes fainter than the VST-ATLAS data that benefits star/galaxy separation accuracy and is still $\approx{1}$ mag deeper when using SDSS Stripe 82 data for $ugri$ photometry (see Section \ref{sec:WHDF_intro}). Since WHDF also has deeper MIR and X-ray imaging, it presents an ideal opportunity to try to optimize our selection methods in this well-observed field. To do this, we start from  the $R$-selected star and galaxy image lists provided on the WHDF webpage \footnote{(http://astro.dur.ac.uk/~nm/pubhtml/herschel/herschel.php)} and match this catalogue to the MIR $3.6$ and $4.5\mu m$ Spitzer SpIES data \citep{Timlin2016} to get an equivalent to $W1$ and $W2$ band photometry. Unless otherwise stated, all magnitudes and colours are corrected for galactic extinction. We next match the Stripe 82 $ugriz$ data to the $R$ image lists of \cite{Metcalfe2001}. We are then able to develop our selection cuts in the WHDF field starting from those described by \cite{Chehade2016} and given in eqs. \ref{eq_chehade_ugr} and \ref{eq_chehade_giw1} above.

\subsubsection{WHDF X-ray \& DESI QSO Population}
\label{sec:WHDF_DESI_Pop}

Firstly, we need to establish the number of known quasars on the WHDF. We consider the X-ray selected sample of WHDF quasars given in Table 2 of \cite{Bielby2012} (see also \citealt{Vallbe2004}), which lists 15 spectroscopically confirmed quasars, their Chandra X-ray fluxes and spectroscopic redshifts. Together with the WHDF morphological and SDSS Stripe 82 photometric properties of these objects, these parameters are all included  in Table \ref{tab:WHDF_bielby_fullinfo} in the Appendix. Of these Chandra X-ray QSOs, 12 are detected at brighter than our target limit of $g<22.5$. These include 10 that are morphologically classified as stellar sources in the WHDF photometric catalogue, and 2 which are classified as extended sources. Additionally, 11 of these 12 quasars are in our `QSO tracer' target redshift range of $0.3<z<2.2$. These 12 confirmed quasars occupy an X-ray-optical overlap area of 214 arcmin$^2$ or 0.0594 deg$^2$, implying  a $16<g<22.5$ 
quasar sky density of $202 \pm 58$ deg$^{-2}$ from the list of \cite{Bielby2012}. Finally, we note that 10 of these 12 X-ray quasars\footnote{WHDFCH099 and WHDFCH113 have $S_X(0.5-10\,{\rm keV})<1\times10^{-14}$ ergs cm$^{-2}$ s$^{-1}$.} are detectable to the nominal eROSITA 0.5-10 keV X-ray flux limit of $1\times10^{-14}$ ergs cm$^{-2}$ s$^{-1}$, corresponding to a sky density of $168 \pm 53$ deg$^{-2}$. 

\begin{table*}
\caption{WHDF completeness and contamination statistics for various
	QSO cut selections  to the ATLAS $g<22.5$ mag limit in all cases. Class `All' means `Stellar' plus `Extended'.  All rows refer to the full redshift range.}
	\centering
	\begin{tabular}{l||c|c|c|c|c|l}
	\hline
    Class	& Cut    & X-ray limit (0.5-10keV)              & Completeness  & Contamination & QSO density & Notes \\
        &        & (ergs cm$^{-2}$ s$^{-1}$)&               &               & (deg$^{-2}$)&       \\
	\hline	
    Stellar &  X-ray & $\ga1\times10^{-15}$     &    11/12$\rightarrow$92\%  &  0/11$\rightarrow$0\%   & 185 &   $>$3$\sigma$ X-ray, 16$<$g$<$22.5 stellar, $<$3$''$\\
    Stellar &  X-ray &     $>1\times10^{-14}$   &    8/12$\rightarrow$67\%  &  0/8$\rightarrow$0\%     & 135 & \\
    Stellar &  DESI  &         -                &    9/12$\rightarrow$75\%  &        -                 & 152 &   \\
    Stellar &  $grW$  &         -                &    11/12$\rightarrow$92\% &  8/19$\rightarrow$42\%   & 185 &   \\
    Stellar &  ugr/UVX   &         -                &     8/12$\rightarrow$67\% &  8/16$\rightarrow$50\%   & 135 &   \\
    \hline
    Extended & X-ray & $\ga1\times10^{-15}$     &    4/4$\rightarrow$100\%   &  6/10$\rightarrow$60\%     & 67 &   $>$3$\sigma$ X-ray, 16$<$g$<$22.5 extended, $<$3$''$\\
    Extended & X-ray &    $>1\times10^{-14}$    &    3/4$\rightarrow$75\%   &  4/7$\rightarrow$57\%    & 51 &    \\
    Extended & DESI  &         -                &    2/4$\rightarrow$50\%   &           -              & 34 &  \\
    Extended & $grW$  &         -                &    3/4$\rightarrow$75\%  &  7/10$\rightarrow$70\%     & 51 & \\
    Extended & ugr/UVX   &         -                &    3/4$\rightarrow$75\%   &  27/30$\rightarrow$90\%   & 51 &  \\	
    \hline
    All      & X-ray & $\ga1\times10^{-15}$     &    15/16$\rightarrow$94\%   &  6/21$\rightarrow$29\%     & 253 &   $>$3$\sigma$ X-ray, 16$<$g$<$22.5, $<$3$''$\\
    All     & X-ray &    $>1\times10^{-14}$    &    11/16$\rightarrow$69\%   &  4/15$\rightarrow$27\%    & 185 &    \\
    All     & DESI  &         -                &    11/16$\rightarrow$69\%   &           -                   & 185  &       \\
    All     & $grW$  &         -                &    14/16$\rightarrow$88\%  &  15/29$\rightarrow$52\%     & 236 & \\
    All      & ugr/UVX   &         -                &   11/16$\rightarrow$69\%   &  35/46$\rightarrow$76\%   & 185 &  \\		
    \hline
	\end{tabular}
	
	\label{tab:WHDF_stats}
\end{table*}

In addition to the Chandra X-ray population of quasars, we also have preliminary DESI Guadalupe internal release data in the WHDF. Here we selected objects that were targeted as QSOs and confirmed spectroscopically as QSOs in these DESI data. These 13 quasars are listed in Table \ref{tab:WHDF_desi_fullinfo}. Note that these data are only preliminary and so future DESI public releases may identify more quasars. But in DESI, there are 11 QSOs to a depth of $g<22.5$, which gives a  density of $185 \pm 56$ deg$^{-2}$, close to the above  X-ray sample of \cite{Bielby2012}. Of these 11 QSOs, 9 are morphologically classified as stellar and 2 as extended. 

There are seven $g<22.5$ QSOs in common between the DESI and the X-ray QSO catalogues. Of the stellar QSOs with $g<22.5$, the DESI and X-ray selected samples  find respectively 2 and 3 QSOs that are undetected by the other technique. Hence we identify a total of 12 stellar QSOs on the WHDF, leading to a stellar QSO density of $202 \pm 58$ deg$^{-2}$. None of the morphologically extended QSOs with $g<22.5$ are in common, meaning there are 4 extended QSOs spectroscopically identified, with 2 in X-ray and 2 in DESI for a total extended QSO sky density of 67 deg$^{-2}$. The total number of $g<22.5$ stellar+extended QSOs on the WHDF is thus 16, corresponding to an overall X-ray+DESI quasar sky density of $269\pm 67$ deg$^{-2}$. 

We note that three out of the four $g<22.5$ DESI quasars missing from Table~\ref{tab:WHDF_bielby_fullinfo} are detected in the X-ray at fainter X-ray fluxes. This increases the overall X-ray completeness from 11/16=69\% at the `eROSITA' $S_X(0.5-10\,{\rm keV})>1\times10^{-14}$ ergs cm$^{-2}$ s$^{-1}$ limit  to 15/16=94\% at the fainter $S_X(0.5-10\,{\rm keV})>1\times10^{-15}$ ergs cm$^{-2}$ s$^{-1}$ `Chandra' limit. These X-ray completenesses can be compared to the overall DESI completeness of 11/16=69\%. Table~\ref{tab:WHDF_stats} provides a  full summary of cut completenesses and contaminations, subdivided by stellar and extended source morphology.

Based on this analysis and subject to the preliminary nature of the DESI internal release, our provisional conclusion is that a joint optical/MIR and `eROSITA' X-ray selection will give an estimated quasar candidate density which is $\approx45$\% higher than simply using the X-ray or DESI optical/MIR selections alone. In particular, we can expect an $\approx45$\% increase in sky density by adding eROSITA X-ray selection to an optical/MIR survey such as DESI to $g<22.5$. Of course, this estimate does not account for any QSOs which may be missed by both techniques.

\subsubsection{WHDF Motivated QSO Cuts}
\label{sec:WHDF_selections}

We now turn our attention to how many of these QSOs are picked up by our photometric selection technqiue, and whether we can optimise this. To test this, we start from the initial $ugri+giW1W2$ photometric cuts, as derived from previous work done by \cite{Chehade2016} and described in Section \ref{sec:2QDESp_cuts}, on the 16 confirmed QSOs. 

\subsubsection{Stellar Cuts}
\label{sec:stellar_cuts}

\begin{figure}
    \centering
	\includegraphics[width=\columnwidth]{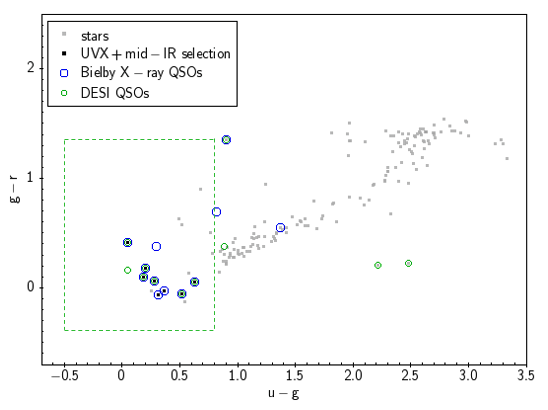} 
	\caption[Colour selections performed on stellar sources in the extended WHDF in the $ugr$ colour space.]{Colour selections performed on stellar sources in the extended WHDF in the $ugr$ colour space. WHDF objects with a stellar morphology are shown in gray. X-ray QSOs from \cite{Bielby2012} are shown as blue circles and QSOs found by DESI are shown as green circles. The $ugr$+$grW1$ ATLAS QSO selections are shown as black points. The green dotted lines denote the ATLAS selection in this colour space. All selections are magnitude limited to $g<22.5$.}
	\label{fig:WHDF_ugr_starcuts}
\end{figure}

We show the 16 WHDF quasars first in the context of the WHDF stellar sources in the same magnitude range in the $u-g:g-r$ plane (see Fig.~\ref{fig:WHDF_ugr_starcuts}). As the WHDF/Stripe 82/SpiES photometry is deeper and less noisy than VST ATLAS/neo 6, we change the $g-r>-1.25$ colour cut of  \cite{Chehade2016} to $g-r>-0.4$ to reflect the reduced contamination in this area. 

\vspace{-0.3cm}
\begin{equation}
\begin{aligned}
    &  -0.5\leq (u-g) \leq 0.8\,\, \\
    &  -0.4\leq (g-r)\leq 1.35
\end{aligned}
\label{eq:whdf_ugr}
\end{equation}

\begin{figure}
    \centering
	\includegraphics[width=\columnwidth]{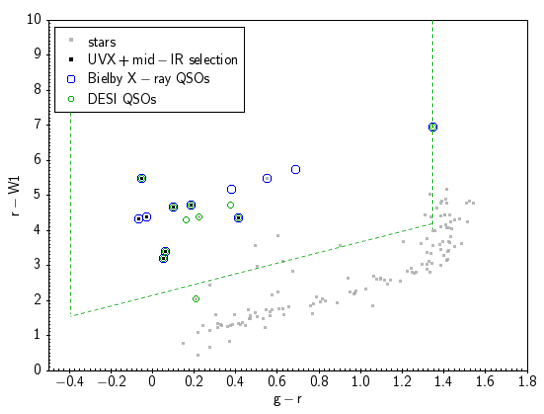} 
	\caption[Colour selections performed on stellar sources in the extended WHDF in the $grW1$ colour space.]{Colour selections performed on stellar sources in the extended WHDF in the $grW1$ colour space. WHDF objects with a stellar morphology are shown in gray. X-ray QSOs from \cite{Bielby2012} are shown as blue circles and QSOs found by DESI are shown as green circles. The $ugr$+$grW1$ ATLAS QSO selections are shown as black points. The green dotted lines denote the ATLAS selections in this colour space. All selections are magnitude limited to $g<22.5$.}
	\label{fig:WHDF_grW1_starcuts}
\end{figure}

We then similarly show the 16 WHDF QSOs in the $g-r:r-W1$ plane\footnote{Here we have moved from the $g-i:i-W1$ plane of \cite{Chehade2016}, for the practical reason that more faint QSOs are detected in ATLAS $r$ rather than $i$. (see Section \ref{sec:final_selections}).} in Fig. \ref{fig:WHDF_grW1_starcuts} where our mid-IR, $grW1$ selections are:

\vspace{-0.3cm}
\begin{equation}
\begin{aligned}
    &  (r-W1) > 1.6*(g-r) + 2.1\,\,  \&\, (i-W1) \leq 8
\end{aligned}
\label{eq:whdf_grw1w2}
\end{equation}

In both figures, the UVX and MIRX ($grW1$) selections are shown as dashed green lines and objects classified as stellar sources are shown in light gray. The stellar locus can be clearly seen in both colour spaces. The X-ray sources are shown as blue circles  and the DESI sources are shown as green circles. Note that none of the 8 extra UVX+$grW1$ candidates (black points) are detected to the $S_X(0.5-10{\rm keV})>1\times10^{-15}$ ergs cm$^{-2}$s$^{-1}$ 3$\sigma$ limit of the Chandra X-ray data. The two X-ray sources and two additional DESI sources that do not overlap with a gray point are morphologically classified as galaxies. 

We show the results of these stellar cuts in Table \ref{tab:WHDF_stats}. We see that the $grW$ is highly successful, selecting 11/12 stellar quasars implying a completeness of 92\% with only  42\% contamination i.e. an efficiency of 58\%. This compares favourably to the other stellar selections eg UVX and X-ray at the brighter `eROSITA' limit both at 67\% completeness. This lower completeness is partly due to both UVX and X-ray being biased against selecting $z>2.2$ quasars, e.g. 2 out of the 4 stellar WHDF quasars missed by UVX have $z>2.2$. One of the other 2 missing in UVX is the  X-ray absorbed, $z=0.79$ quasar, WHDFCH0044, which may explain its red $u-g=0.89$ colour. The other is WHDFCH055 at $z=0.74$ which is much redder at $u-g=1.37$ but shows little evidence of X-ray absorption.
However, UVX still has a competitively low contamination rate for stellar quasars at 50\% compared to 42\% for $grW$ and we shall see that UVX still has a role to play when the imaging data is less deep and the star-galaxy separation is less accurate.

\subsubsection{Extended Source Cuts}
\label{extended_source_cuts}

As 5 of the 15 confirmed QSOs from \cite{Bielby2012} and a further 3 DESI QSOs (or 7 in total accounting for one overlap) are morphologically classified as extended sources (galaxies) in the WHDF catalogue, we perform our colour selections on extended sources as well. Down to $g<22.5$, even the star/galaxy separation in the WHDF data is not entirely reliable, so our decision to include this selection will be further justified when looking at images with lower S/N as in the VST ATLAS survey. At this $g<22.5$ limit, 2 extended QSOs are found by \cite{Bielby2012} and 2 by DESI.

\begin{figure}
    \centering
	\includegraphics[width=\columnwidth]{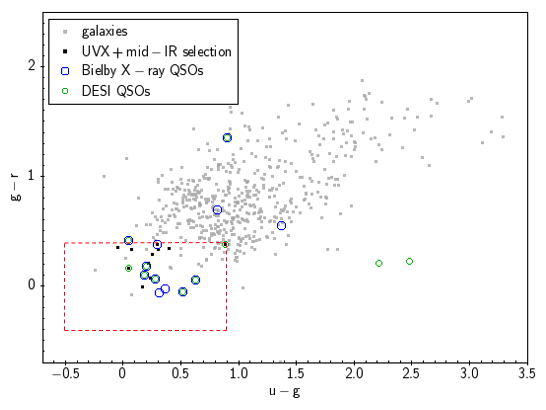} 
	\caption[Colour selections performed on extended sources in the WHDF in the $ugr$ colour space.]{Colour selections performed on extended sources in the WHDF in the $ugr$ colour space. WHDF extended sources (galaxies) are shown in gray.  X-ray QSOs from \cite{Bielby2012} are shown in blue and QSOs found by DESI are shown in green. There are 2 extended QSOs from \citep{Bielby2012} and 2 extended QSOs from DESI which can be seen to have extended counterparts. The $ugr$+$grW1$ ATLAS QSO selections for extended sources are shown as black points. The red dotted lines denote the ATLAS selection in this colour space. All selections are magnitude limited to $g<22.5$. Note that, although difficult to see on the plot, the QSO at u-g=1.37 does not have an extended counterpart.}
	\label{fig:WHDF_ugr_galcuts}
\end{figure}

\begin{figure}
    \centering
	\includegraphics[width=\columnwidth]{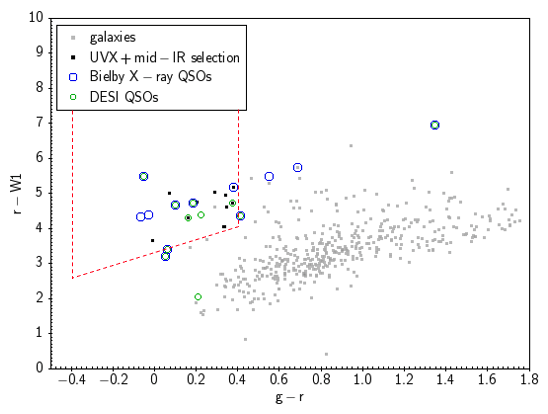} 
	\caption[Colour selections performed on extended sources in the WHDF in the $grW1$ colour space.]{Colour selections performed on extended sources in the WHDF in the $grW1$ colour space. WHDF extended sources (galaxies) are shown in gray. X-ray QSOs from \cite{Bielby2012} are shown in blue and QSOs found by DESI are shown in green. There are therefore 2 extended QSOs  from \cite{Bielby2012} and 2 extended QSOs from DESI. The $ugr$+$grW1$ ATLAS QSO selections for extended sources are shown as black points. The red dotted lines denote the ATLAS selection in this colour space. All selections are magnitude limited to $g<22.5$. }
	\label{fig:WHDF_giW1_galcuts}
\end{figure}

The suggested cuts, shown in Figs.~\ref{fig:WHDF_ugr_galcuts} and~\ref{fig:WHDF_giW1_galcuts}, are aimed at minimising galaxy contamination while retaining possible QSOs that have been classified as galaxies.\footnote{Note that the $z=2.68$ DESI QSO (ID 8222 in Table \ref{tab:WHDF_desi_fullinfo}) is found to be a double object in deep WHDF H-band imaging with $0.''9$ seeing.}. In these figures, the X-ray sources and DESI sources which are {\it not} overlapping with gray points are morphologically classified as stellar. 

These restricted $ugr$ cuts for extended sources are as follows:

\vspace{-0.3cm}
\begin{equation}
\begin{aligned}
    & -0.5\leq (u-g) \leq 0.9\,\, \\
    &  -0.4\leq (g-r) \leq 0.4\,\,  \\
\end{aligned}
\label{eq:gal_ugr}
\end{equation}

\noindent The restricted mid-IR $grW$ cuts are:

\vspace{-0.3cm}
\begin{equation}
\begin{aligned}
    &  (r-W1)\geq 1.6*(g-r)+3.3 \,\, 
\end{aligned}
\label{eq:gal_giw1}
\end{equation}

The two  X-ray QSOs with $g<22.5$  from \cite{Bielby2012} that are morphologically classified as extended sources (WHDFCH20 and WHDFCH110) have redshifts of $z=0.95$ and $z=0.82$. Visual inspection suggests that WHDFCH110 might be slightly elongated and that WHDFCH020 might overlap a faint galaxy in the $r$-band. The two DESI QSOs classified as galaxies with $g<22.5$ are WHDF 8222 at $z=2.68$ and WHDF 3081 at $z=1.31$. WHDF 3081 is also found to be a relatively bright X-ray source, WHDFCH052, listed by \cite{Vallbe2004} but not by \cite{Bielby2012}. WHDF 8222 is similarly listed as a fainter X-ray source by \cite{Vallbe2004} - see Table \ref{tab:WHDF_desi_fullinfo}. We have already noted that the $z=2.68$ QSO is a double object (in $H$-band) and  probably lensed. The $z=1.31$ QSO appears to be interacting with a pair of very red compact sources at $\approx3''$ distance. We conclude on the basis of these 4 extended QSOs that they are mostly not mis-classifications and should be included in our QSO sample. This is supported by  other, fainter, $g>22.5$ QSOs, WHDFCH007, WHDFCH008 that are also classed as galaxies on an WHDF HST $i$ image  \citep{Shanks2021}. Finally, WHDFCH048 that also has $g>22.5$ and is classed as a galaxy, although outside the HST $i$ frame, appears to be interacting with 2 other objects within $\approx3''$, again justifying its extended classification.

\subsection{WHDF Selection Summary and Conclusions}
\label{sec:WHDF_conclusions}

To summarise, we have tested our photometric selections in the extended WHDF Chandra X-ray overlap area of 214 arcmin$^2$ or 0.0594 deg$^2$. The main results from the WHDF analysis as tabulated in Table \ref{tab:WHDF_stats} are :-

1) A complete census of the broad lined QSO population in the WHDF to $g<22.5$ using X-ray, $ugr$, $grW$ and also DESI results reveals a total confirmed QSO sky density of $269\pm67$ deg$^{-2}$ at $16<g<22.5$. From their Luminosity Function (LF) model, \cite{PD2016} estimate 196 deg$^{-2}$ at this limit, again in good statistical agreement with the $269\pm67$ deg$^{-2}$  found in the WHDF. These authors also predict 143 deg$^{-2}$ at $z<2.2$, within $\approx1\sigma$ of the $202\pm58$ deg$^{-2}$ found on the WHDF. They also predict 53 deg$^{-2}$ at $z>2.2$, again in good agreement with the $67\pm17$ deg$^{-2}$ found in the WHDF. We also note that 25\% of all WHDF QSOs to $g<22.5$ were morphologically classed as galaxies/extended in the R-band, a sky density of $67\pm34$ deg$^{-2}$.

2) The X-ray QSOs have a sky density of $253\pm65$ deg$^{-2}$ with $g<22.5$ to the faint Chandra limit and  $185\pm56$ deg$^{-2}$  with $g<22.5$ to $S_X(0.5-10\,{\rm keV})>1\times10^{-14}$ ergs cm$^{-2}$ s$^{-1}$, approximately the `eROSITA' limit. Of these $g<22.5$ X-ray QSOs, $\approx20$\% were classed as extended.

3) From  the DESI optical-MIR selection  a total sky density of $185\pm56$ deg$^{-2}$ $g<22.5$ QSOs were found,  of which $101\pm41$ deg$^{-2}$ were detected as X-ray QSOs at the eROSITA X-ray limit and  $84\pm38$ deg$^{-2}$ were undetected at this limit. Again 20\% were classed as extended and 80\% were stellar.

4) We conclude that  neither X-ray (at the brighter `eROSITA' limit) nor the preliminary DESI data produce complete stellar QSO samples, both missing $\approx30$\% of stellar QSOs to $g<22.5$.  Similarly  X-ray and DESI miss $\approx50-60$\% of extended QSOs. So, they give a stellar QSO sky density of 135-152 deg$^{-2}$  and  extended QSO sky densities of 34-51 deg$^{-2}$, leading to a sky density for both of 185 deg$^{-2}$. Given the total WHDF QSO sky density of $269\pm67$ deg$^{-2}$ this means that both have a similar overall completeness of $\approx70$\%, implying that an eROSITA X-ray survey will add $\approx40-45$\% to a $g<22.5$ optical/MIR QSO sky density. We also note that X-ray selected, stellar sub-samples have essentially zero contamination, much less than any other selection method. 

5) For DESI QSOs with $g<22.5$, 4/11 have $z>2.2$, implying a sky density of $\approx$67 deg$^{-2}$ and 7/11 having $z<2.2$ for  a sky density of $\approx$118 deg$^{-2}$.  X-ray selection is always more skewed towards lower redshifts (e.g. \citealt{Boyle1994}), with none here at the brighter `eROSITA' limit having $z>2.2$. But note that at the fainter $S_X(0.5-10\,{\rm keV})>1\times10^{-15}$ ergs cm$^{-2} $s$^{-1}$ limit,  three of these four $z>2.2$ QSOs are ultimately  also detected in X-rays.

6) In principle, a stellar $grW$ cut should select $\approx90$\% of the QSOs for a sky density of  $\approx185$ deg$^{-2}$ while suffering $\approx38$\% contamination. The stellar X-ray selection to the `eROSITA' limit  produces 67\% completeness, for a sky density of  $\approx$135 deg$^{-2}$ with zero contamination, at least when matched to a $g<22.5$ star sample. The UVX technique produces similar $\approx67$\% completeness with only slightly lower $\approx33$\% contamination. For extended sources, the $grW$, UVX and `eROSITA' X-ray selections all achieve  75\% completeness which is only bettered by the 100\% completeness of the  faint `Chandra' X-ray selection. The X-ray selections have the lowest contamination and the UVX selection the highest.

7) Thus focusing first on optimising QSO selection in the stellar samples, and assuming no X-ray data is available, $grW$ seems the most promising base for selection giving higher completeness and lower contamination than $ugr$. For the 20-25\% of QSOs classed as extended, although the $grW$ and $ugr$ completenesses are the same, the contamination is lower for $grW1$ than $ugr$.

So the MIRX cuts generally perform better than UVX when the optical photometry is as deep as in the WHDF and when the MIR photometry is as deep as in the SpIES survey. But we again emphasise that these results apply only in the best quality data as is available in the WHDF. In particular, we shall see below that at SDSS or ATLAS depths with no X-ray data yet available, the $ugr$ selection still has an important role to play alongside $grW$ in selecting $z<2.2$ and $z>2.2$ QSO samples at $g<22.5$.

\section{VST-ATLAS QSO Selection}
\label{sec:final_selections}

Based on previous experience with VST ATLAS and the WHDF analysis above we now describe our QSO selection using the current VST ATLAS data. As stated in Section \ref{sec:ATLAS}, our VST-ATLAS data set was updated from previous work. Therefore, we begin by noting that we have improved the star/galaxy separation by adding to the standard separation in $g$ an additional selection in the $g_{\rm Kron}-g_{A3}:g$ plane\footnote{$g_{A3}$ is the $g$ magnitude measured within a $1''$ radius aperture.} to account for seeing variation in interchip gaps covered by only one of the two stacked images \citep{Shanks2015}. Here we used the relations $g_{\rm Kron}-g_{A3}>(0.5g_{A3}-0.864)$ for $g<19.78$ and $g_{\rm Kron}-g_{A3}>0.125$ for $g>19.78$ to select extra stars from amongst the objects initially classified as galaxies. To increase the depth of our survey, we also introduce a seeing weighted combination of the ATLAS $u-$band magnitude and the Chilean $u-$band extension program.

As the ATLAS data is noisier than the data available in the WHDF, we have to adjust slightly the selections used there to decrease contamination. This can be seen in the ATLAS $u-g:g-r$ selection in Eq. \ref{eq:ugr4} which more closely follows the wider ATLAS stellar locus. We shall see that basic $grW$ cuts in ATLAS give a high contamination, leading to candidate densities of up to $\approx$400 deg$^{-2}$ caused by galaxy contamination. As we do not yet have full X-ray coverage of VST-ATLAS, we therefore pursue joint MIRX and UVX selections which seemed to reduce contamination by $\approx10$\% even in the high quality WHDF data (see Section \ref{sec:WHDF_conclusions}). Therefore, instead of selecting either only the $grW$ MIRX candidates OR the $ugr$ UVX candidates, we shall combine these with the aim of providing a high priority (called Priority 1 for the rest of this paper) $16<g<22.5$ QSO candidate catalogue, dominated by $z<2.2$ QSOs because of the inclusion of the UVX cuts. 

But first, following \cite{Croom2009} and further motivated by experience with deeper KiDS $ugri$ data in the GAMA G09 field (see Eltvedt et al 2022, in prep.), we apply a cut to remove White Dwarf stars that would otherwise contaminate our UVX selection.
\medskip

\noindent Base selection with White Dwarf cut: 

\vspace{-0.3cm}
\begin{equation}
\begin{aligned}
& 16< g \leq 22.5\,\, \&\,\, -0.4\leq (g-r) \leq 1.1 \,\, \& \\
& not\ [0.44(g-r)-0.17<(r-i)<0.44(g-r)-0.02 \\
& \&\,\,(g-r)<-0.05)]
\end{aligned}
\label{eq:wd_2}
\end{equation}

The selections using our UVX and our mid-IR colour cuts are then defined as follows:
\medskip

\noindent ATLAS UVX selections:
 
\vspace{-0.3cm}
\begin{equation}
\begin{aligned}
& -0.5 \leq (u-g) \leq 0.65\ || \\
& (u-g)<0.65\,\, \&\,\, (g-r) \leq -0.9(u-g)+0.8
\end{aligned}
\label{eq:ugr4}
\end{equation}

\noindent Mid-IR, known hereafter as $grW$, selections:

\vspace{-0.3cm}
\begin{equation}
    (r-W1) > 0.75(g-r) + 2.1\,\, \&\,\,     (W1-W2)>0.4
\label{eq:grw1}
\end{equation}

\noindent The last $W1-W2$ cut is only performed on objects which are found using the mid-IR selections with a detection in $W2$. If they have no detection in $W2$, only the mid-IR selections featuring W1 are applied.
\medskip

\begin{figure}
	\centering
	\includegraphics[width=\columnwidth]{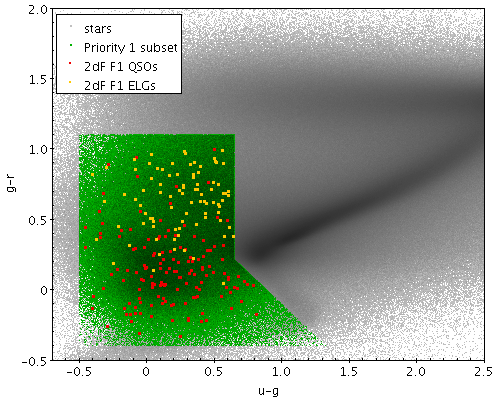}
	\caption[Final ATLAS ugr selection]{The final $u-g:g-r$ selection for VST-ATLAS Priority 1 QSO candidates. We also show the placement of the 2dF F1 objects which were observed in this ugr colour space. Objects which have been identified as QSOs are shown in red, NELGs are shown in yellow. ATLAS stars are shown in gray and the Priority 1 sample is shown in green. }
	\label{fig:final_ugr}
\end{figure}

\begin{figure}
	\centering
	\includegraphics[width=\columnwidth]{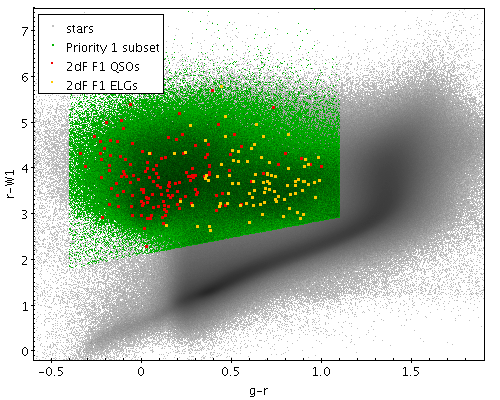}
	\caption[Final ATLAS $grW1$ selection]{The final $g-r:r-W1$ selection for VST-ATLAS priority 1 QSO candidates. We also show the placement of the 2dF F1 objects which were observed in this $grW1$ colour space. Objects which have been identified as QSOs are shown in red, NELGs are shown in yellow. ATLAS stars are shown in gray and the Priority 1 sample is shown in green. }
	\label{fig:final_grW1}
\end{figure}

\begin{figure*}
	\centering
\includegraphics[width=\textwidth]{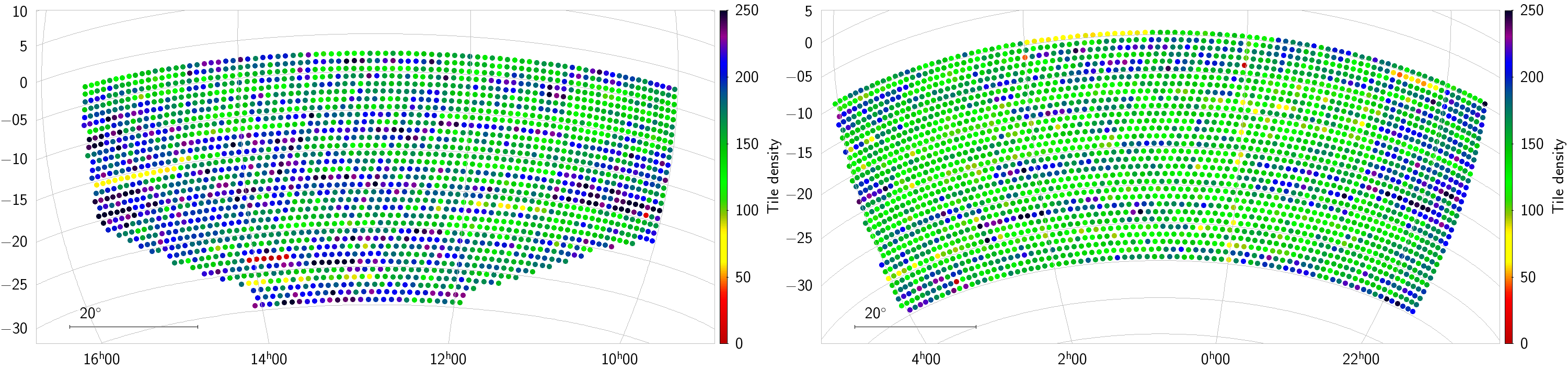}
	\caption[Tile density of QSO candidates in the NGC and SGC using the "Priority 1" selection]{VST-ATLAS NGC and SGC tile density (deg$^{-2}$) of $ugr$ \& $grW$ Priority 1 QSO candidates.}
	\label{fig:qso_priority1_tiledensity}
\end{figure*}

The main $grW$ and UVX selections can be seen in Figs. \ref{fig:final_ugr} and \ref{fig:final_grW1}. Here we show the ATLAS stellar objects in gray, with the stellar loci clearly visible. The candidates selected through our Priority 1 sample are shown in green.  The resulting tile density of QSO candidates targeted through the Priority 1 selections can be seen in Fig.~\ref{fig:qso_priority1_tiledensity} for the NGC and SGC.

After the first star-galaxy separation step of  QSO selection,  we  noticed significant gradients in the sky density of stars, particularly in the NGC and, to a lesser extent, in the  SGC and these persisted into the final QSO samples such as the Priority 1 selection shown in the Fig.~\ref{fig:qso_priority1_tiledensity}.
The fluctuations within field concatenations are of order of  $\sim \pm 18\%$ (full range) in the NGC and $\sim \pm 9\%$ in the SGC, bigger than expected from Poisson fluctuations. The extra contributions mainly come from  increased galaxy contamination in fields with poorer $g$-band seeing,  residual $\sim20\%$ tile incompleteness in Chilean $u$-band data  where only ATLAS u-band data is available and stellar features like the Sagittarius stream which covers the NW corner of the ATLAS SGC area. This  feature also caused a similar gradient in the DESI target catalogue and this had to be removed prior to  the QSO angular clustering analysis of  \cite{Chaussidon2022}.

After a first round of observing on the 2dF instrument at AAT (see Section \ref{sec:2dF_comparison}), we found  that the main contaminants of the Priority 1 selection are compact Narrow Emission Line Galaxies (NELGs), with these source accounting for  about $25\%$ of the contamination. Figs. \ref{fig:final_ugr} and \ref{fig:final_grW1} show spectroscopically confirmed QSOs in red and NELGs in yellow. The latter seem to cluster in a cloud centred at $g-r\approx0.7$ and $r-W1\approx3.5$ Therefore, we define a further cut to be optionally excluded from this Priority 1 subset in order to reduce this galaxy contamination. This NELG `exclusion zone' is defined as:

\vspace{-0.3cm}
\begin{equation}
    g>22\,\, \&\,\, (r-W1) < 2.5(g-r)+2.5
\label{eq:nelg}
\end{equation}

\begin{figure*}
	\centering
\includegraphics[width=\textwidth]{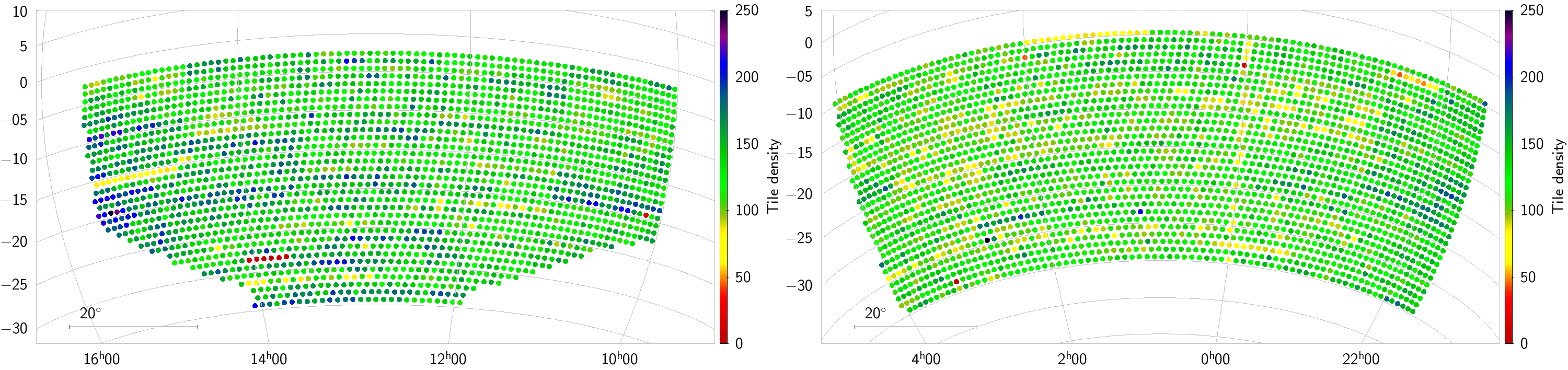}
	\caption[Tile density of QSO candidates in the NGC and SGC using the "Priority 1" selection as well as an NELG cut]{VST-ATLAS NGC and SGC tile density (deg$^{-2}$) of $ugr$ \& $grW1W2$ QSO candidates with  an additional selection to remove NELGs.}
	\label{fig:qso_prioritynoelg_tiledensity}
\end{figure*}

\noindent This cuts down the QSO candidate sky densities by 41 and 31 deg$^{-2}$ in the NGC and SGC. The resulting QSO tile density across the sky from this selection which reduces the NELG contamination  can be seen in Fig.~\ref{fig:qso_prioritynoelg_tiledensity} for the NGC and SGC. 

\begin{figure*}
	\centering
\includegraphics[width=\textwidth]{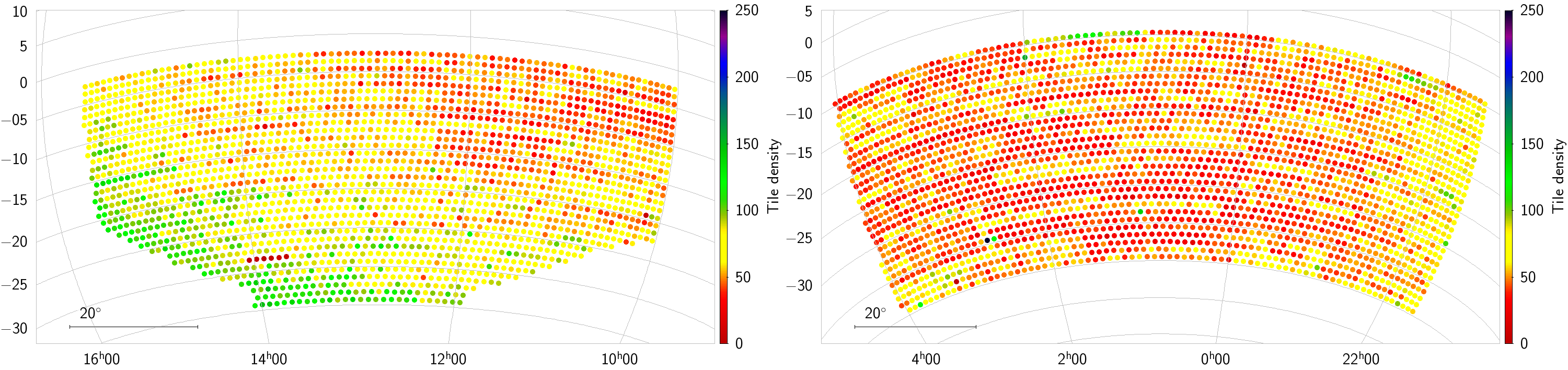}
	\caption[Tile density of QSO candidates in the NGC and SGC using the IRX selection without UVX]{VST-ATLAS NGC and SGC tile density (deg$^{-2}$) of MIRX \& non-UVX candidates to target higher redshift objects. Note the significant gradient to higher sky densities towards lower Galactic latitudes (i.e. RA$\approx15$h, Dec=-20 deg).}
	\label{fig:qso_grW1W2nougr_tiledensity}
\end{figure*}

We also define a selection to target higher redshift, $z>2.2$ objects. For this selection, we target objects found in the MIRX selections that are not detected through our UVX selection, also requiring a detection in W2. The tile density of candidates for this selection are seen in Figure~\ref{fig:qso_grW1W2nougr_tiledensity}. This selection, defined below in Eq.~\ref{eq:grw_nonuvx}, is referred to as `$grW$ non-UVX' throughout the rest of the paper.

\vspace{-0.3cm}
\begin{equation}
\begin{aligned}
    & 16< g \leq 22.5\,\, \&\,\, -0.4\leq (g-r) \leq 1.1 \,\, \& \\
    & (r-W1) > 0.75(g-r) + 2.1\,\, \&\,\,     (W1-W2)>0.4\,\, \& \\
    & ![0.44(g-r)-0.17<(r-i)<0.44(g-r)-0.02 \\
    & \&\,\,(g-r)<-0.05)]\,\, \& \\
    & !(-0.5 \leq (u-g) \leq 0.65\ || \\
    & (u-g)<0.65\,\, \&\,\, (g-r) \leq -0.9(u-g)+0.8)
\end{aligned}
\label{eq:grw_nonuvx}
\end{equation}
\\

\begin{figure*}
	\centering
\includegraphics[width=\textwidth]{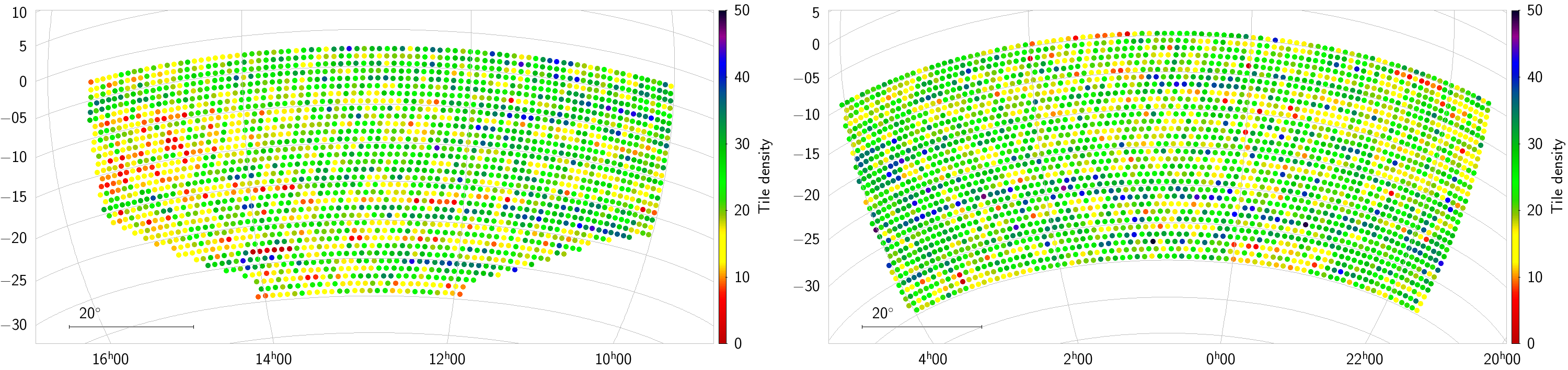}
	\caption[Tile density of QSO candidates in the NGC and SGC that are classified as galaxies in the g-band]{VST-ATLAS NGC and SGC tile density (deg$^{-2}$) of QSO candidates that are classified as galaxies in the g-band.}
	\label{fig:qso_gal_tiledensity}
\end{figure*}

Finally, we define a selection for QSOs that we believe have been (mis-)classified as galaxies. For this selection, we start with the previously defined extended source cuts (as outlined in Section~\ref{extended_source_cuts}). We adjust the $u-g$ cut in the same way as the stellar selection. We also introduce the $W1-W2$ requirement to decrease contamination. The final extended source selection is shown below in Eq.\ref{eq:gal_final}. The tile density of QSO candidates targeted with this selection is seen in Fig.~\ref{fig:qso_gal_tiledensity}.

\vspace{-0.3cm}
\begin{equation}
\begin{aligned}
    & -0.5\leq (u-g) \leq 0.65\,\, \&\\
    &  -0.4\leq (g-r) \leq 0.4\,\, \&  \\
    &  (r-W1)\geq 1.6*(g-r)+3.3 \,\, \&\\
    &  (W1-W2)>0.4\,\,  
\end{aligned}
\label{eq:gal_final}
\end{equation}

The overall sky densities of these selections are shown in Table \ref{tab:ATLAS_final_numbercounts}. The NGC has an $\approx24$\% higher candidate density than the SGC in the Priority 1, $grW$ non-UVX  and total cases, the only exception being the galaxy cut. Since the NGC is at lower galactic latitudes than the SGC it is likely that this is caused by higher stellar contamination. However, since the main contaminants are expected to be NELGs a more complicated explanation might be needed such as the higher stellar density causing more galaxy-star overlaps that disrupt the $grW$ stellar rejection via colour contamination. Otherwise the candidate densities are reasonably homogeneous in Figs. \ref{fig:qso_priority1_tiledensity} - \ref{fig:qso_gal_tiledensity} with the main exception being the NGC high redshift selection where an increasing candidate density toward lower galactic latitudes is seen in Fig. \ref{fig:qso_grW1W2nougr_tiledensity}. We shall see that similar results apply once we split into $z>2.2$ and $z<2.2$ samples using photometric redshifts in Section \ref{sec:ANNZ}.

\begin{table*}
    \caption{Number counts and sky densities for the colour selections applied to the full VST-ATLAS footprint. Totals  in column 8 are sum of columns  4, 6, and 7.}
    \begin{tabular}{ p{2cm}||p{1.5cm}|p{1.5cm}|p{2cm}|p{2cm}|p{1.5cm}|p{1.5cm}|p{1.5cm} }
      \hline
      Sky Area & UVX       & $grW$    & UVX \& $grW$ & UVX \& $grW$ & $grW$ \& & extended & total \\
               & selection & selection & "Priority 1"              & with NELG cut &non-UVX        & cuts   & candidates \\
      \hline
      NGC (2034 deg$^2$) & 1,128,470 & 985,294 & 395,459 & 312,296 & 154,723 & 49,556 & 599,738 \\
      NGC (deg$^{-2}$) & 554.8 & 484.4 & 194.4 & 153.5 & 76 & 24.4 & 294.9 deg$^{-2}$ \\
      SGC (2706 deg$^2$) & 910,719 & 834,994 & 422,731 & 339,194 & 142,204 & 64,315 & 629,250 \\
      SGC (deg$^{-2}$) & 336.6 & 308.6 & 156.2 & 125.3 & 52.6 & 23.77 & 232.5 deg$^{-2}$ \\
    \hline
      Total (4740 deg$^2$) & 2,039,189 & 1,820,288 & 818,190 & 651,490 & 296,927 & 113,871 & 1,228,988 \\
    Total (deg$^{-2}$) & 430.2 & 384.0 & 172.6 & 137.4 & 62.6 & 24.0 & 259.3 deg$^{-2}$ \\
     \hline
    \end{tabular}

    \label{tab:ATLAS_final_numbercounts}
\end{table*}

\section{Spectroscopic Completeness and Efficiency of the VST-ATLAS QSO Selection}
\label{sec:spectroscopic_completeness}

We utilise photometric DESI QSO candidate target catalogues along with spectroscopic results which will be released in DESI DR1 (see \cite{Chaussidon2022_targetselection}, \cite{Alexander2022}, \cite{Myers2022} for information on DESI target selection and data quality validation), as well as our own spectroscopic results from 2dF in order to test the completeness and efficiency of our ATLAS QSO candidate selection to our faint $g<22.5$ limit. We also similarly utilise 2QZ, 2QDES, and eBOSS, which are completed spectroscopic surveys and also have large areas of overlap with VST ATLAS to test our final ATLAS QSO catalogue down to their respective $g<20.8$, $g<22$, and $g<21.9$ magnitude limits. Taken together, these analyses provide a reasonably complete picture of the completenes and efficiency of our full  ATLAS QSO catalogue. 

\subsection{DESI Comparison}
\label{sec:DESI_comparison}

The latest DESI internal data release, Guadalupe, covers a large area of the DESI footprint which includes some $\approx$144 deg$^2$ overlap with VST ATLAS (see Fig. \ref{fig:DESI_ATLAS_area}). In addition to the Guadalupe release, we also utilize the DESI quasar candidate catalogue/quasar targets in this area, which were chosen using DECaLS Legacy Survey DR9 data \citep{Yeche2020}, to form a more complete comparison of our quasar candidate selections. In order to test first the accuracy of the ATLAS photometry down to $g<22.5$,  we look initially at an $\approx$8.5 deg$^2$ sub-area of the larger $\approx$144 deg$^2$ overlap with DESI targets in the NGC centred around RA=14$^{\rm h}$08$^{\rm m}$00.0$^{\rm s}$, Dec$=-4\degree$. This area encompasses approximately one DESI rosette, which has 5000 fibre positions, including sky fibres.

\subsubsection{DESI-ATLAS photometric comparison}

\begin{figure*}
	\centering
l	\includegraphics[width=\textwidth]{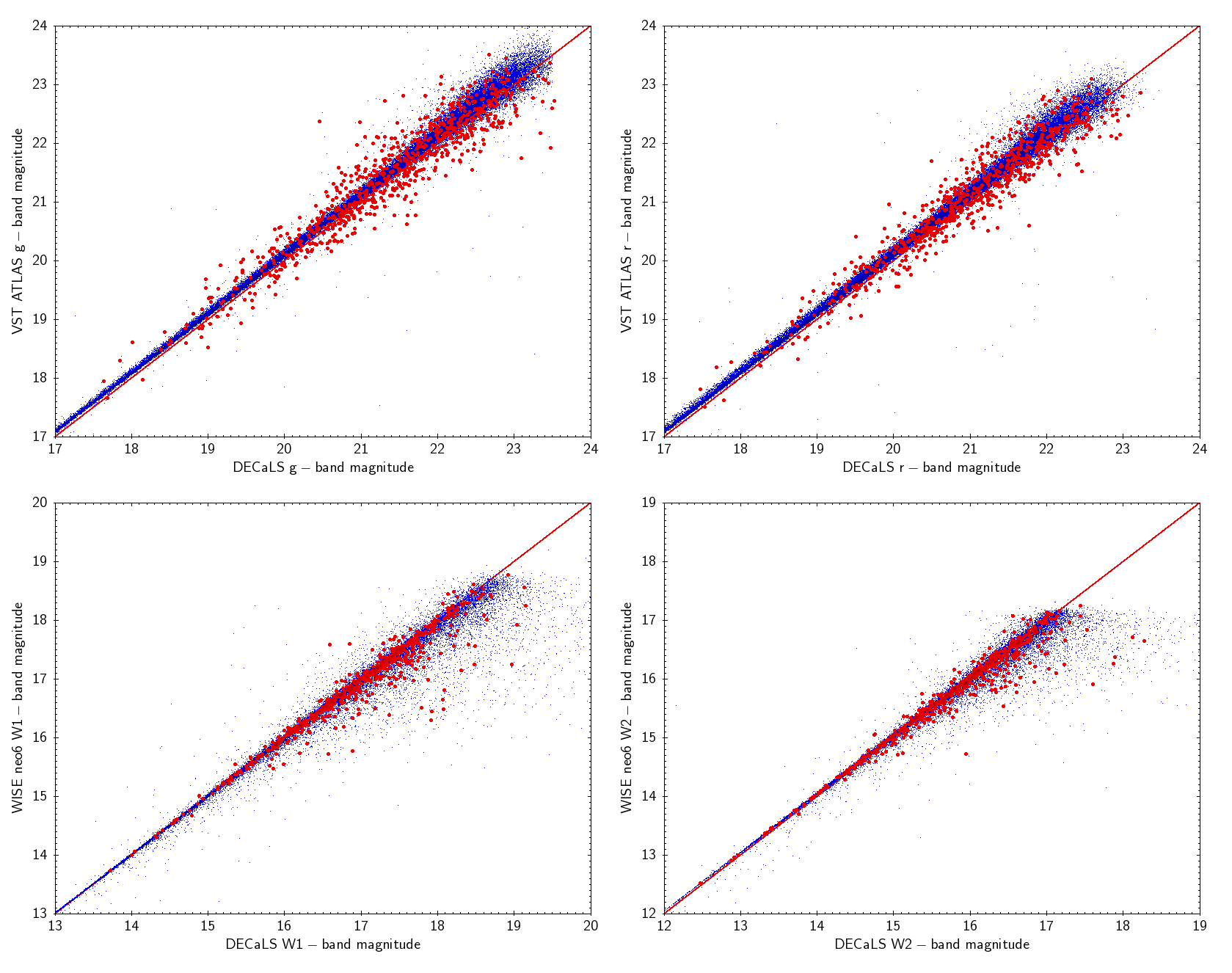}
	\caption[DESI DECaLS DR9 vs. VST ATLAS in the $g-$, $r-$, $W1-$ and $W2-$bands]{DESI DECaLS DR9 vs. VST ATLAS in the $g-$, $r-$bands and DESI DECaLS DR9 v WISE (neo6) in the W1- and W2-bands, in a sample $\sim8.5$ deg$^2$ area. The red points represent known QSOs and the blue points represent objects classified as stars.}
	\label{fig:grW1W2_DECaLS_ATLAS_comparison}
\end{figure*}

We first check the photometric quality of our VST ATLAS data by matching the raw $g$-band, $r$-band, and $W1$-band data to the DESI targets in the 8.5 deg$^2$ sub-area of Fig. \ref{fig:DESI_ATLAS_area}. The results can be seen in Fig. \ref{fig:grW1W2_DECaLS_ATLAS_comparison}. Generally we see good agreement between the depth of the ATLAS aperture 3 $g$ and $r$ stellar photometry compared to DECaLS, as well as the WISE neo6 data vs. DECaLS. However, we see that the QSO candidates show a larger scatter than the general stars, especially in $g$ and $r$, and particularly at brighter magnitudes.

Comparisons between both SDSS and ATLAS data and SDSS and DESI data show a similarly large scatter for quasar candidates. This suggests that this excess scatter, particularly at bright magnitudes, is dominated by quasar variability caused by the significant epoch difference between these 3 datasets. Indeed, even at $21.5<g<22.5$, the scatter in stars remains at only the $\pm0.05$mag level implying that our ATLAS photometry remains accurate at and perhaps even beyond our $g=22.5$ limit. The result in $r$ is similar with a scatter of only $\pm0.05$ mag measured in the range $22<r<23$ mag.\footnote{We also note the presence of small offsets in both $g$ and $r$ - these are of order a few hundredths of a magnitude, and are due in part to small colour terms between the DECaLS and ATLAS passbands.}

\subsubsection{DESI - ATLAS Target Overlap}

Now we have determined that our data quality is comparable down to our limit of $g<22.5$, we check target overlap in the $\approx$144 deg$^2$ area of overlap between DESI observations and the NGC of VST-ATLAS, seen in Fig. \ref{fig:DESI_ATLAS_area}. There are 37,306 ATLAS quasar candidates in this area, giving a 259 deg$^{-2}$ sky density. In the same area there are 50,016 DESI QSO candidates. These were selected through a combination of photometric colour cuts and a Random Forest code (see e.g. \citealt{Yeche2020}), and are limited to $r<23$. Of these, 34,106 lie within our $16<g<22.5$ range, giving a 237 deg$^{-2}$ target density. When we match these DESI targets to our full VST ATLAS catalogue (prior to making any QSO selections), we get a match of 29,897  objects, with 4,209 (=34,106-29,897 or 12\%) being unmatched in ATLAS. If we now perform a match between the DESI targets and our total ATLAS QSO candidate selection, using a $1''$ matching radius, we find 17,673 overlapping objects. This includes our full $ugr+grW$ selection, the $grW$ non-$UVX$ selection, as well as our `galaxy' selection. Therefore, our ATLAS quasar selections are missing 12,224 (=29,897-17,673=41\%) of the 29,897 QSO candidates selected by DESI and that are available in ATLAS, giving an ATLAS `target completeness' relative to DESI of 59\%. Of these 12,224 objects, $58\%$ are morphologically classified as stars, $41\%$ are classified as galaxies, $48\%$ do not have a detection in $W1$, and $41\%$ were removed due to the base $gri$ White Dwarf cut. Comparing our VST ATLAS selections individually, the $ugr+grW$, Priority 1, cut  gives us 24,676 candidates in this area, of which 12,765 are in common with DESI QSO candidates. The non-UVX selection  has 9,296 candidates, with 3,703 in common. Finally, the extended source selection has 3,334 candidates, with 1,652 in common.

\begin{table*}
\caption{NGC ATLAS-DESI overlap test of the various ATLAS QSO selections. The selections in column 1 are described in Section~\ref{sec:final_selections}. `Priority 1' is comprised of objects found in both 'star $grW$' and `star UVX'. `Star total' includes the `priority 1' objects in addition to the `star $grW$ non-UVX' candidates. Column 2 shows the sky density of QSO candidates based on each selection. Column 3 shows the completeness of the ATLAS selections with regards to the DESI spectroscopically confirmed QSOs across the full redshift range. This is then split between the $z<2.2$ and $z>2.2$ completeness in columns 4 and 5.
}
    \begin{tabular}{ l||c|c|c|c}
      \hline
ATLAS Subset   & Sky density   & Completeness      & $z<2.2$ & $z>2.2$ \\
               &  (deg$^{-2}$) &  (\%)             &    (\%) &    (\%)  \\
      \hline
\hline
Stars $16<g<22.5$&             &                   &           &    \\
\hline
star $grW$ &56656/144=393 &8735/10107=86.4&6803/7656=88.9&1932/2451=78.8\\
star $grW$ (W2 required) &23534/144=163&7870/10107=77.9&6373/7656=83.2&1497/2451=51.1\\
star UVX & 71844/144=499&7071/10107=70.0&6051/7656=79.0& 1020/2451=41.6\\
Priority 1 & 24676/144=171 &6708/10107=66.4 &5818/7656=76.0& 890/2451=36.3\\
star $grW$ non-UVX& 9296/144=65  &1707/10107=16.9 &918/7656=12.0& 789/2451=32.2\\
star total & 33972/144=236  &8415/10107=83.3 &6736/7656=88.0&1679/2451=68.5\\
      \hline
Galaxies $16<g<21.9$&          &                 &            &          \\
\hline
extended cuts &  3334/144=23&   912/3267=27.9 &  858/2675=32.1 & 54/592=9.12 \\
\hline
Total          &               &                 &              &\\
     \hline
               & 37306/144=259 & 9327/13374=69.7 &7594/10331=73.5 & 1733/3043=57.0\\
     \hline
    \end{tabular}

    \label{tab:ATLAS_DESI_u_W2}
\end{table*}

\subsubsection{DESI-ATLAS spectroscopic comparison}

The DESI collaboration started commissioning their main spectroscopic survey at the start of 2021. We shall be using spectra from the Guadalupe internal data release and we emphasise that all results reported here must be regarded as preliminary because they were taken from a snapshot of the DESI spectroscopic  catalogue that may be incomplete in terms of exposure time, area coverage, etc. We use the QSO catalogues described by ~\cite{Chaussidon2022_targetselection} and Alexander et al (2022), constraining our sample to `dark' and `bright' main programs, and again focusing on the $\approx$144 deg$^2$ overlap with VST ATLAS as indicated in Fig. \ref{fig:DESI_ATLAS_area}. This provides at least an initial estimate of DESI completeness and efficiency for the purpose of evaluating the same parameters for  VST ATLAS. We emphasise that at this stage not all the DESI candidates have been spectroscopically targeted, and the DESI results may change when their full coverage and exposure times are achieved.

There are a total of 17,716 spectroscopically confirmed DESI QSOs in the 144 deg$^2$ area, although only 14,302 lie in the range $16 < g < 22.5$, and not all of these were originally targeted as QSO candidates. Overall, only 17,553 of the 34,106 DESI $g<22.5$ QSO candidates have so far been observed, with 13,128 of these confirmed to be QSOs, so 14,302-13,128=1,174 were presumably selected through other targeting programs, such as ELGs (\citealt{Raichoor2022}), LRGs (\citealt{Zhou2022}) or bright galaxies (\citealt{Hahn2022}).  The 13,128/17,553=74.7\% DESI QSO fraction is higher than the 70-71\% success rate suggested by \cite{Chaussidon2022_targetselection} and \cite{Alexander2022}, probably due our application of a $g<22.5$ limit rather than the full $r<23$ DESI limit.

The full VST-ATLAS data overlaps with 13,374 of the 17,716 quasars. Out of these 13,374 possible quasars that were available to $g_{\rm ATLAS}<22.5$, our QSO selections picked up 9,327 objects, composed of 7,594 out of the 10,331 DESI QSOs at $z<2.2$ and 1,733 out of the 3,043 DESI QSOs at $z>2.2$ (see Table \ref{tab:ATLAS_DESI_u_W2}). Thus the overall ATLAS completeness at $g<22.5$ relative to DESI is 70\% in this field, with 74\% at $z<2.2$ and 57\% at $z>2.2$. 

We note that there remains advantage to be gained from including the ATLAS $u$-band in our selection as well as $grW$. Using the stellar $grW$ selection in Eq. \ref{eq:grw1} would result in an overall sky density of 393 deg$^{-2}$ and this reduces to 171 deg$^{-2}$ by combining with stellar UVX selection to give  Priority 1  in Table \ref{tab:ATLAS_DESI_u_W2}. Although including $grW$ non-UVX increases this by 65 deg$^{-2}$  to 236 deg$^{-2}$, this represents an $\approx40$\% reduction in candidate density. However, if we use $grW$ (which already requires $W1-W2>0.4$ for those objects with $W2$) with the added demand that only objects with a measured $W2$ are included then the density reduces to 163 deg$^{-2}$. But the total stellar selection with the $u$ band still achieves a completeness with respect to DESI of 83\% compared to 78\% with $grW$ (W2 required). The completeness advantage is slightly  bigger for the $z>2.2$ sample than for the $z<2.2$ sample (see Table \ref{tab:ATLAS_DESI_u_W2}). 

There remains $393-236\approx$157 deg$^{-2}$ $grW$ candidates that are not included in either the UVX or non-UVX samples. These could still be treated as lower priority candidates in a spectroscopic survey.

We finally recall from \cite{Chehade2016} that at the depth of the AllWISE W1 and W2 data used by these authors,  the MIRX candidates only reached $g\approx20.5$ mag whereas with neo6 the depth reached is $g\approx22$ mag in W1 and $g\approx21.5$ mag in W2. With further NEOWISE exposure time the W1, W2 depth reached will be highly competitive with UVX so that in the cases at least where deep, high resolution $griz$ photometry is available then the $u$ data may not be required. The $griz$ photometry that is available  in the DES area satisfies these conditions and so the 4CRS survey may not require the availability of $u$ data to reach the same depths as in VST ATLAS.

\subsubsection{DESI comparison conclusions}

To summarize, the DESI QSO candidate sky density at $g<22.5$, over the full redshift range is 237 deg$^{-2}$. Using a success rate of 74.7\%, based on the 13128/17553 spectroscopically confirmed objects, and assuming the observed objects are a random selection from the candidate list, we can estimate that the DESI Guadalupe release currently has a $g<22.5$ quasar sky density of 178 deg$^{-2}$. We again emphasise that this result may ultimately change due to the preliminary nature of the DESI internal data release used here. The DESI QSO density over their full magnitude and redshift range is quoted in \cite{Chaussidon2022_targetselection} as $>200$ deg$^{-2}$ with an efficiency of $\sim71\%$ based on their main selection. 

If we extrapolate these results to $r<23$ using a canonical $N\propto10^{0.3m}$ we find that the sky density  rises from  237 deg$^{-2}$ to 335 deg$^{-2}$ compared to 310 deg$^{-2}$ quoted by \cite{Chaussidon2022_targetselection} (see also \citealt{Alexander2022}). Similarly, the 178 deg$^{-2}$ QSO sky density we find at $g<22.5$ increases to 251 deg$^{-2}$ compared to the $>200$ deg$^{-2}$ indicated by \cite{Chaussidon2022_targetselection}. Given that the DESI numbers are restricted to $z>0.9$ whereas ours apply to $z>0.5$, we regard these numbers as being in reasonable agreement. 

Our VST ATLAS QSO candidate sky density in the DESI overlap area at $g<22.5$ is 259 deg$^{-2}$. Based on the spectroscopic completeness relative to DESI (see above), we can extrapolate that the ATLAS confirmed $g<22.5$ quasar sky density is $0.7\times178=125$ deg$^{-2}$. Therefore, the ATLAS efficiency at $g<22.5$ and all $z$ is $125/259=48.2$\%. 
However, when we look at the efficiency of our targets that were observed by DESI, 10595 of the ATLAS targets in the overlap area were observed, of which 9327 were confirmed to be QSOs. Therefore, we have a $9327/10595=88\%$ efficiency of observed targets. This higher efficiency than the 75\%  and 48\% DESI and ATLAS efficiencies noted above, is likely due to  jointly selected targets naturally having lower contamination rates than either individual selection. 

Finally, we can determine that if we assume DESI Guadalupe is already complete in the area we have used and that Guadalupe samples $z<2.2$ and $z>2.2$ targets fairly, then the DESI sky density at $z<2.2$ will be $178\times 10331/13374=137$ deg$^{-2}$ and at $z>2.2$ it will be $178\times 3043/13374=41$ deg$^{-2}$. The ATLAS sky density at $z<2.2$ will then be $0.74\times137=102$ deg$^{-2}$ and $0.57\times41=24$ deg$^{-2}$ for $z>2.2$.

At $g<22.5$, for all redshifts, our ATLAS selection is missing confirmed DESI QSOs. The ATLAS $grW1$ bands all seem comparatively deep enough. In the specific case of $W1$, we tested this by swapping the DECaLS DR9 $W1$ for the neo6 $W1$ band and  finding that this resulted in  little change to the selected candidates. Additionally, the missing QSOs are located in same place in the $gri$, $grW1$ and $ugr$ colour spaces as the confirmed quasars. The main problem seems to be in $W2$ with 2058/13374= 15\% of DESI-ATLAS confirmed quasars missing in neo6 $W2$. 

Improved ATLAS morphological star/galaxy separation might reduce our galaxy contamination but, as we have seen, quasars can be correctly classed as extended and NELGs exist that are compact and stellar like. Thus until deeper $W2$ data becomes available we require to use the joint MIRX and UVX selection to limit galaxy contamination while maintaining a high completeness. 

We also note that the VST ATLAS 125 deg$^{-2}$ quasar sky density at $g<22.5$ is a lower limit because there are likely to be extra quasars in the ATLAS candidate list that did not appear in the DESI list. These extra ATLAS quasars could be those that had varied to be brighter than $g<22.5$ at the ATLAS epoch while being dimmer than the DESI limit ($r<23$) at the DESI epoch. We shall see there is some evidence for this effect in the 2dF tests of ATLAS cuts in Section \ref{sec:2dF_comparison} below.

\subsection{2dF Comparison}
\label{sec:2dF_comparison}

We were further able to test our selection through observing runs using the 2dF instrument with the AAOmega spectrograph \citep{Sharp2006} at the Anglo-Australian Telescope (AAT) in February-April, 2021 (see Table \ref{tab:2df}). Two fields were observed, NGC-F1 and NGC-F2/NGC-F2A. The 580V and 385R gratings were used with the 5700\AA~dichroic. Both fields were run first with targets from our standard ATLAS quasar UVX+$grW$ selection. The NGC-F2A observation then prioritised the $grW$ non-UVX and the extended source selections. Most data was obtained for NGC-F1 with 4.75 hrs of observations and it is clear that such an exposure time is needed to get as high as $\approx67$\% spectroscopic identifications, given the average observing conditions that were experienced. The exposure time for the NGC-F2 observation was less than half that of NGC-F1 resulting in only 54\% spectral identifications achieved.

After the first 2dF run on NGC-F1 we noted that there was significant contamination by Narrow Emission Line Galaxies (NELGs) with a sky density of $\approx$50 deg$^{-2}$. So for the F2 observation we applied a further $gri$ stellar cut to reduce this contamination (see Section \ref{sec:final_selections}, Eq. \ref{eq:nelg}). This did reduce the NELG contamination but also contributed to the lower F2 quasar sky densities (see Table \ref{tab:2df}) and so this further NELG cut is not advised when trying to maximise quasar sky densities.

\begin{table*}
    \caption{ATLAS  Fields observed by 2dF. For the NGC-F2A data in the bottom row, the blue and red arm of the spectra were reduced and analyzed separately. Here we show what fields were observed, for how long, the seeing on each field, and what percentage of the data we were able to make spectroscopic QSO IDs on.}
    \begin{tabular}{p{1.5cm}||p{1.25cm}|p{1.25cm}|p{1.5cm}|p{2.5cm}|p{1.25cm}|p{1.25cm}|p{1.25cm}|p{2cm}}
    \hline
    Field   &RA(deg)&Dec(deg)&  Date     &  Exposure         & Seeing   & IDs    & Total Exp. &  Comments\\
    \hline
    NGC-F1  & 196.9 & -16.0 &  18/2/2021 & 1$\times$30mins+4$\times$20mins & $2.''1$  &        &  $--$      &      \\
    NGC-F1  & 196.9 & -16.0 &  09/3/2021 & 1$\times$25mins+2$\times$30mins & $1.''4$  &        &  $--$      &       \\
    NGC-F1  & 196.9 & -16.0 &  15/3/2021 & 3$\times$30mins          & $4.''0$  & 66.8\% &  4.75hr    &       \\
    NGC-F2  & 211.6 & -16.0 &  09/3/2021 & 18.3+25+15.3mins  & $1.''5$  & 60.2\% &  $--$      &        \\
    NGC-F2  & 211.6 & -16.0 &  15/3/2021 & 3$\times$25mins          & $4.''0$  & 35 (54)\%&  2.25hr  &   \\ 
    NGC-F2A & 211.6 & -16.0 &  07/4/2021 & 4$\times$20mins          & $2.''5$  & 43\%   &  1.33hr    & Moon\\
    \hline
    \end{tabular}
 	\label{tab:2df}
\end{table*}

\begin{table*}
    \caption{2dF NGC-F1 and NGC-F2/F2A 2dF+AAOmega  spectroscopic identifications. The $z<2.2$ and $z>2.2$ columns describe spectroscopically confirmed QSOs. The percentages in columns 4 and 6 show the efficiency of our selection at both redshift ranges. \dag~ implies that an extra NELG cut was used.}
    \begin{tabular}{p{2.5cm}||p{2.35cm}|p{0.75cm}|p{1.3cm}|p{1.3cm}|p{1.3cm}|p{1.3cm}|p{0.75cm}|p{0.75cm}|p{0.75cm}}
    \hline
    Field & Candidates & Fibred & $z<2.2$ & $z<2.2$ &  $z>2.2$ & $z>2.2$ &  NELGs      & Stars      & No ID   \\
     & & & QSOs & QSOs & QSOs & QSOs & & & \\
     &  (\#)  &   (\#)   &(N/Percent) &  (deg$^{-2}$) &  (N/Percent)  &(deg$^{-2}$) &(deg$^{-2}$)&(deg$^{-2}$)&(deg$^{-2}$)  \\
    \hline
    NGC-F1 UVX       & 561 & 352 & 203/57.7\%    & 107.8& 28/8.0\%      &   14.9      & 52.5       & 2.4        &  17.3   \\
    NGC-F2 UVX$^\dag$& 486 & 347 & 154/44.4\%    & 71.9 & 24/4.9\%      &   11.2      & 19.6       & 4.7        &  54.6     \\
    NGC-F2A non-UVX  &187(g$<$21.1)& 182 & 5/2.7\% & 1.7   & 36/19.8\%     &   12.0      &  4.8       & 5.1        &  33.3  \\
    NGC-F2A galaxies & 102 & 65 & 2/0.0\%         &0.7   &  9/11.2\%     &    3.0      &  4.9       & 1.2        &  15.2    \\
    NGC-F2A NELG cut & 127 & 102 & 0/0.0\%         &0.0   & 19/11.2\%     &    7.9      &  5.8       & 0.8        &  16.2    \\
    \hline	
    Total & 187+62+34=283deg$^{-2}$ & 599 & 208  & $>$110.2 &   64          &   $>$29.9   &$>$62.2     &  $>$8.7    &  50.6  \\  
    \hline
    \end{tabular}
	\label{tab:ngcf1f2a}
\end{table*}

\begin{table*}
    \caption{Completeness and efficiency of the VST ATLAS QSO candidates based on DESI and 2dF, from Tables \ref{tab:ATLAS_DESI_u_W2} and \ref{tab:ngcf1f2a}.}
    \begin{tabular}{p{0.6cm}||p{1.2cm}|p{1.2cm}|p{0.8cm}|p{0.8cm}|| p{0.6cm}|  p{0.6cm}|p{1.2cm}|p{1.2cm}|p{0.8cm}| p{0.8cm}  }
      \hline
   Survey &ATLAS & ATLAS &ATLAS & ATLAS& Pri 1& Pri 1 &Non-UVX &Non-UVX& Ext.& Ext.\\
       & candidates& QSOs& Comp.& Eff.        &Comp. & Eff. &Comp.& Eff. & Comp.& Eff. \\
      \hline
DESI &259 deg$^{-2}$&125 deg$^{-2}$ & 70\% & 48.3\% & 66\% & 52\% & 17\% & 35\% & 28\%  & 53\%\\

2dF  & 283 deg$^{-2}$& 140 deg$^{-2}$ & N/A  & 50\% & N/A & 66\% & N/A & 22\%& N/A &11\% \\
     \hline
    \end{tabular}
    \label{tab:DESI_2dF_summary}
\end{table*}

In what follows, we therefore focus on the combination of the NGC-F1 UVX+$grW$, priority 1, selection and the F2A non-UVX and extended source selection. In NGC-F1, we have $561$ priority 1 QSO candidates. Of these $561$ candidates, $352$ were fibred. After analyzing the resulting spectra in MARZ \citep{Hinton2016}, we find that $231$ of these are identified as having $QOP=3$ or 4 redshfts (where $QOP$ is the MARZ spectral quality parameter with $QOP= 3,4$ implying redshift qualities `good' and `excellent'). This is $65.6\%$ of our target list, which gives us 122.7 deg$^{-2}$ QSOs when normalized to the full number of targets at the same priority level in the field. We find 88 NELGs, giving us a galaxy contamination of $25\%$, or 46.8 deg$^{-2}$. There are 4 stars, which results in an $8.2\%$ stellar contamination, or 15.4 deg$^{-2}$. Finally, there are $29$ objects which have no clear ID, a rate of $8.2\%$, or the equivalent of 15.4 deg$^{-2}$ in our priority candidate subset. Furthermore, of the 231 spectroscopically identified QSOs, we find 203 at $z<2.2$, giving a sky density of 107.8 deg$^{-2}$ in our target redshift range, and $28$ QSOs at $z>2.2$ giving a sky density of 14.9 deg$^{-2}$.

In NGC-F2 lower QSO ($QOP=3$ or 4) sky densities were found  with only 71.9 deg$^{-2}$ at $z<2.2$  and 11.2 deg$^{-2}$ at $z>2.2$ identified in the $ugr+grW$ selection, compared to 107.8 and 14.9 deg$^{-2}$ with the the same selection in the NGC-F1 field.

So, as summarised in Table~\ref{tab:ngcf1f2a}, the AAT 2dF observations of NGC-F1 and NGC-F2A suggest that by combining the F1 priority 1 and the F2A non-UVX and extended source selections, achieves a $z<2.2$ QSO sky density of 110 deg$^{-2}$ and a $z>2.2$ sky density of 30 deg$^{-2}$ for a total sky density of 140 deg$^{-2}$. With a combined candidate density of 283 deg$^{-2}$, this implies an ATLAS efficiency of 140/283=50\%. These and the other ATLAS efficiencies are summarised in Tables \ref{tab:ngcf1f2a} and \ref{tab:DESI_2dF_summary}. We see there is reasonable agreement between the results found in the DESI area and the 2dF field. These two tests complement each other with the DESI area giving lower limits on confirmed quasar sky densities from ATLAS because DESI itself may not be complete. The 2dF efficiencies will be upper limits especially at $z<2.2$ because of the $g<21.1$ limit that had to be used due to a lack of 2dF fibres for the NGC-F2A $grW$ \& non-UVX  sample (termed `NGC F2A non-UVX' in Table \ref{tab:ngcf1f2a}). 

\begin{table*}
    \caption[Completeness in the SGC] {VST-ATLAS completeness in the SGC based on spectroscopically confirmed QSOs from 2QZ, 2QDES, and eBOSS. Confirmed QSOs-stellar and Confirmed QSOs-exten. refer to the number of confirmed QSOs that are picked up as stars and galaxies, respectively, in the VST-ATLAS catalogue prior to making any QSO selection. }
    \begin{tabular}{p{2cm}||p{1.5cm}|p{1.5cm}|p{1.5cm}|p{1.5cm}||p{1.5cm}|p{1.5cm}||p{2cm}  }
      \hline
      Survey& Confirmed    & Overlap    & Overlap $grW$  & Completeness& Confirmed     & Overlap   & Completeness  \\
            &QSOs-stellar  & Priority 1 & non-UVX &  (Star total)  & QSOs-exten.   &gal cut  & (Stellar+Extended) \\
      \hline
      2QZ ($g<20.8$)     & 10179 & 9372 & 544 & $97.4\%$ & 1672 & 939 & $91.6\%$  \\
      2QDES ($g\la22$)   & 2258 & 1962 & 130  & $92.6\%$   & 232  & 105 & $88.2\%$  \\
      eBOSS($g$$<2$$1.8$)& 1495 & 1148 & 270  & $94.8\%$ & 221  & 78  & $87.2\%$  \\
     \hline
    \end{tabular}
    \label{tab:SGC_final_completeness_efficiency}
\end{table*}

\begin{table*}
    \caption[Completeness in the NGC]{VST-ATLAS completeness in the NGC based on spectroscopically confirmed QSOs from 2QZ, 2QDES, and eBOSS. Confirmed QSOs-stellar and Confirmed QSOs-exten. refer to the number of confirmed QSOs that are picked up as stars and galaxies, respectively, in the VST-ATLAS catalogue prior to making any QSO selection. }
    \begin{tabular}{p{2cm}|p{1.5cm}|p{1.5cm}|p{1.5cm}|p{1.5cm}||p{1.5cm}|p{1.5cm}||p{2cm}  }
      \hline
      Survey& Confirmed    & Overlap    & Overlap $grW$  & Completeness& Confirmed     & Overlap   & Completeness  \\
            &QSOs-stellar  & Priority 1 & non-UVX &  (Star total)  & QSOs-exten.   &gal cut  & (Stellar+Extended) \\
      \hline
      2QZ ($g<20.8$)       & 1337 & 1216 & 88 & $97.5\%$ & 188 & 106 & $92.5\%$ \\
      2QDES ($g\la22$)     & 4175 & 3417 & 204 & $86.7\%$ & 134 & 64 & $85.5\%$  \\
      eBOSS($g$$<2$$1.8$)  & 1855 & 1230 & 399 & $87.8\%$ & 282 & 86 & $80.3\%$ \\
     \hline
    \end{tabular}
    \label{tab:NGC_final_completeness_efficiency}
\end{table*}

\subsection{2QZ, 2QDES, eBOSS Comparison}

We also utilize previously completed spectroscopic surveys to assess further the completeness and efficiency of our VST-ATLAS quasar selections. The completeness for each selection, compared to spectroscopically confirmed QSOs from 2QZ, 2QDES, and eBOSS, can be seen in Table ~\ref{tab:SGC_final_completeness_efficiency} for the SGC and Table ~\ref{tab:NGC_final_completeness_efficiency} for the NGC through the individual `Overlap' columns as well as the final `star total' and `stellar+extended' completeness columns. The confirmed QSOs-stellar column refers to the total number of confirmed QSOs in each respective survey that, when matched to the full VST ATLAS survey, are classified as stars through our star/galaxy classifications. The confirmed QSOs-exten. column is the number of confirmed QSOs that are classified as a galaxy in our classifications. 

The main result here is that in the brightest 2QZ sample the ATLAS stellar (star total) selections are producing $\approx97$\% completeness. These completenesses reduce for objects classed as extended but only to $\approx92$\%. It is not clear why this is the case but the poorer completeness for extended objects might be explained if they contained more lensed double quasars that were prone to higher variability, for example. We note that 1672/11851 = 14\% of 2QZ SGC quasars are classed as extended, with a similar fraction in the NGC, further justifies our inclusion of extended sources in our selections. The lower completenesses in 2QDES and eBOSS are mainly due to their fainter magnitude limits, possibly allied to higher variability if they are gravitationally lensed.  The eBOSS sample is mainly a sub-sample of the DESI quasars in areas near the Dec$\approx0$ deg Equatorial regions.

\subsection{Spectroscopic analysis conclusions}


 
Through our comparisons of DESI and ATLAS, our own observations from 2dF, and comparisons with 2QZ, 2QDESp, and eBOSS, we are able to estimate the completeness and efficiency of our VST ATLAS QSO candidates. The main results of these analyses, as shown in Tables~\ref{tab:ngcf1f2a}, \ref{tab:DESI_2dF_summary}, \ref{tab:SGC_final_completeness_efficiency}, and \ref{tab:NGC_final_completeness_efficiency} are:





\begin{enumerate}
\item From DESI comparisons, we estimate the overall VST ATLAS QSO completeness at 70\%. At brighter magnitudes we see higher completenesses in the range 88-97\% from comparisons with 2QZ, 2QDESp, and eBOSS.\\

\item From DESI and 2dF  comparisons, we estimate the VST ATLAS QSO efficiency in the range 48-50\%. We thus estimate the ATLAS true QSO sky density to be in the range of 125-140 deg$^{-2}$ for our full redshift range, with 102-110 deg$^{-2}$ at $z<2.2$ and 24-30 deg$^{-2}$ at $z>2.2$.
\end{enumerate}

\section{Final ATLAS QSO Catalogue}
\label{sec:final_catalogue}

As summarised in Table ~\ref{tab:ATLAS_final_numbercounts}, our final Priority 1 quasar candidate counts in the NGC give us a sky density of 194 deg$^{-2}$, and a sky density of 156 deg$^{-2}$ in the SGC. The colour selections performed on galaxies give an additional candidate sky density of 24 deg$^{-2}$ in both the NGC and SGC. The mid-IR, $grW$ non-UVX candidates give us a sky density of 76 deg$^{-2}$, and a sky density of 53 deg$^{-2}$ in the SGC. 

\begin{figure*}
	\centering
\includegraphics[width=\textwidth]{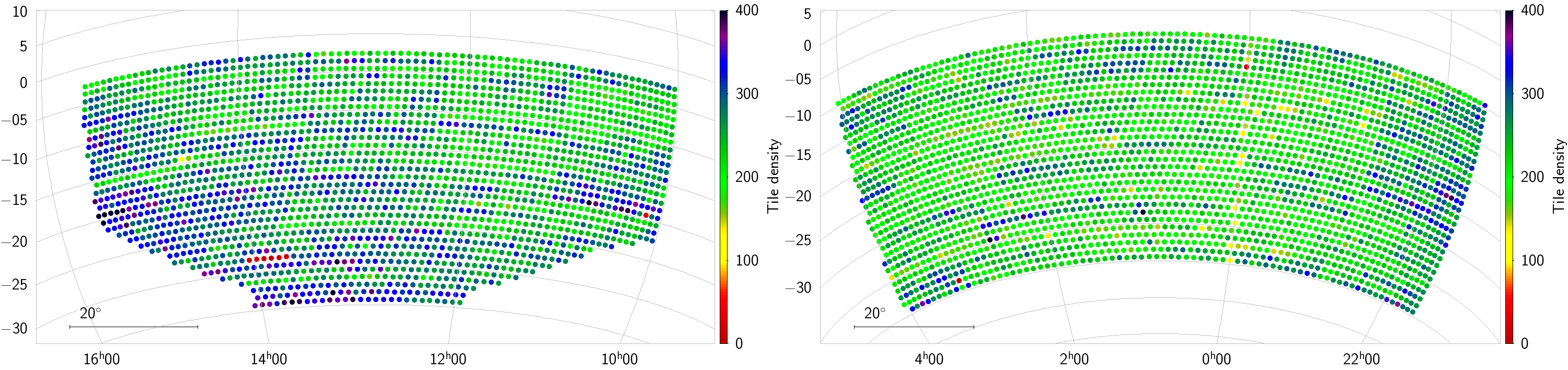}
	\caption[Tile density of total QSO candidates in the NGC and SGC ]{VST-ATLAS tile density (deg$^{-2}$) of the total number of QSO candidates in the NGC and SGC using our full Priority 1 + $grW$\&non-UVX + extended selections.}
	\label{fig:qso_total_tiledensity}
\end{figure*}

Combining the NGC and SGC gives a sky density of 173 deg$^{-2}$ Priority 1 candidates, plus 63 deg$^{-2}$ non-UVX candidates, plus 24 deg$^{-2}$ with the additional extended source selections. The 65\% higher candidate sky densities seen in the NGC for the UVX selection is probably due to the lower Galactic latitudes covered by the NGC, causing higher amounts of star contamination. In the Priority 1 sample the NGC is only 24\% higher because of the intrinsically low stellar contamination in combining the UVX and $grW$ selections, which allows the more isotropic quasar distribution to dominate. As can be seen in Fig.~\ref{fig:qso_total_tiledensity}, the quasar candidate sky density across the NGC and SGC is relatively uniform, barring some striping most likely due to sky conditions such as seeing and sky brightness.

\subsection{n(g)}

\begin{figure*}
	\centering
	\includegraphics[width=\textwidth]{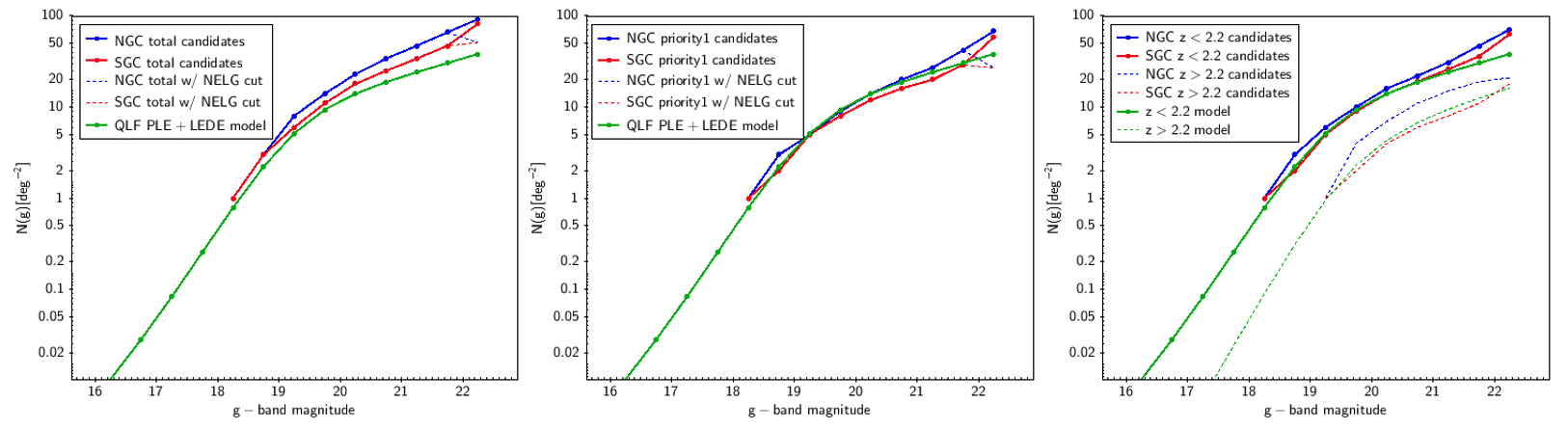}
	\caption[Final QSO g-band number counts compared to the QLF PLE$+$LEDE model]{The observed VST ATLAS QSO candidate NGC and SGC number-magnitude counts  compared to the PLE$+$LEDE QLF model predictions of \protect\cite{PD2016}. (a) The full NGC+SGC VST-ATLAS QSO candidate sky densities as a function of $g-$band magnitude are shown in blue and red. The predicted QSO number counts from the QLF PLE$+$LEDE model in the $0<z<4$ redshift range are shown in green. (b) Same as (a) for the Priority 1 counts compared to the $0<z<2.2$ model. (c) Same as (a) for $z_{\rm photo}<2.2$ (blue and red solid lines) and $z_{\rm photo}>2.2$ (blue and red dashed lines). 
	}
	\label{fig:Full_QSO_gband}
\end{figure*}

In Fig.~\ref{fig:Full_QSO_gband}, we now compare our candidate QSO number counts to the pure luminosity-function plus luminosity and density evolution (PLE+LEDE) model \cite{PD2016}, based on the quasar luminosity function (QLF) measured in eBOSS in the redshift range $0.68<z<4.0$. This QLF data is fit by a double power-law model, with a linear pure luminosity-function for redshifts of $z<2.2$ combined with a luminosity and density evolution model at $z>2.2$. This new QLF is then used to predict the expected quasar number counts in order to optimize the fibre targeting for DESI to their limit of  $r\approx23$. They updated their selection algorithm based on the time variability of quasar fluxes in SDSS Stripe 82. From Table 6 of \cite{PD2016}, we take the expected quasar number counts, which are presented in bins of $\Delta g=0.5$ mag and $\Delta z=1$ for the magnitude range of $16<g<22.5$ and the redshift range of $0<z<3$.

These expected number counts are shown   in Fig.~\ref{fig:Full_QSO_gband}, first for their full $0<z<3$ redshift range. They predict a quasar candidate sky density of 196 deg$^{-2}$ over this redshift range at $g<22.5$, consistent with the $269\pm67$ deg$^{-2}$ we estimated from the deep WHDF data in Section ~\ref{sec:WHDF_DESI_Pop}.   This predicted  sky density can  be compared to our full Priority 1+`$grW$+non-UVX'+extended quasar selection which gives a candidate sky density of 259 deg$^{-2}$ at $g<22.5$, 32\% higher than the \cite{PD2016} QLF PLE$+$LEDE $0<z<3$ model, due mostly to contamination which is highest in the non-UVX and extended source cuts. Using their Table 6, we also estimate a rough quasar candidate sky density in the $0<z<2.2$ range, more appropriate for comparison with our Priority 1 sample. Therefore, compared to  the 195 deg$^{-2}$ candidates at $0<z<3$, we find $\approx$143 deg$^{-2}$ in the redshift range of $0<z<2.2$ compared to 173 deg$^{-2}$ in our Priority 1 sample and 137 deg$^{-2}$ if the NELG cut is also made (see Table ~\ref{tab:ATLAS_final_numbercounts}). This agreement to within $\approx4$\% of model (143 deg$^{-2}$) vs. Priority 1 with NELG cut (137 deg$^{-2}$)  is reasonably consistent with the low contamination rate for the Priority 1 sample found in the NGC-F1 2dF data  when the NELG cut is applied (see Table~\ref{tab:ngcf1f2a}).

As in Section \ref{sec:final_selections}, we note that the NGC sky density at 295 deg$^{-2}$ is significantly ($\approx26.7$\%) higher than the SGC at 232.8 deg$^{-2}$. Now it is likely that this is simply due to higher contamination in the NGC, especially with the known  NELG contamination of the raw Priority 1 sample and the increased contamination of the non-UVX and extended source cuts. However, the NGC count remaining high relative to the SGC over the large $18<g<22$ range seen in Fig.~\ref{fig:Full_QSO_gband} (a) is somewhat surprising given the high efficiency/low contamination of QSO selection at bright, $g<21$, magnitudes.

To investigate this effect further, we again restrict ourselves to just the Priority 1 candidates, that in the main have $z<2.2$ due to the inclusion of the UVX criterion. They are therefore also more comparable to counts to brighter limits selected only by UVX, such as 2QZ, 2SLAQ and SDSS. For example, at the 2QZ limit of $g<20.8$ the sky density at $z<2.2$ is known to be $\approx35$ deg$^{-2}$, rising to $\approx40$ deg$^{-2}$ after 2QZ completeness correction (e.g. \citealt{Croom2009}). But the main reason we focus on the Priority 1 candidates is their high efficiency/low contamination which facilitates model and NGC vs SGC count comparison. So in Fig.~\ref{fig:Full_QSO_gband} (b) the Priority 1 NGC and SGC $n(g)$ counts  are compared to the QLF PLE$+$LEDE model, now over the redshift range of $0.5<z<2.2$. Here, again we see that the NGC sky density at $g<22.5$ remains higher than the SGC count, now by 24.4\% (194.4 vs 156.2 deg$^{-2}$ - see Table \ref{tab:ATLAS_final_numbercounts}).  We also note that the NGC count remains consistently higher than the SGC count over the $18<g<22$ range. So we now limit the Priority 1 sample at $g<20.8$ where we expect a true QSO sky density of $\approx40$ deg$^{-2}$. We find that the NGC sky density is 46.8 deg$^{-2}$ whereas the SGC sky density is 40.2 deg$^{-2}$, so the NGC Priority 1 count is 16.3\% higher than the SGC. To find what is causing this excess contamination in the NGC, we can look back at the NGC-F1 2dF observations at this same limit. In this field the Priority  1 candidate density to $g<20.8$ was 44.1 deg$^{-2}$, so similar to the NGC average of 46.8 deg$^{-2}$, within error. At $g<20.8$, NGC F1 sky densities were QSO 39.3 deg$^{-2}$, NELG 1.6 deg$^{-2}$, Stars/WD 0.5 deg$^{-2}$ and non-IDs 2.6 deg$^{-2}$. So assuming that the non-IDs are not QSOs this implies only  $\approx$11\% contamination in this typical NGC field. So this contamination barely takes us to the level of the SGC which would require $\approx16$\% \ contamination in the Priority 1 NGC sample. Since it is likely that an SGC field observed for as long as NGC-F1 would also have similar contamination, it is not clear that increased contamination in the NGC does explain its increased sky density relative to the SGC. Deeper 2dF data in an SGC field to determine the amount of contamination there is required to resolve this question of this apparent NGC-SGC anisotropy. We return to these issues at the end of Section~\ref{sec:ANNZ}.

\section{ANNz2 Photometric Redshift Estimation}
\label{sec:ANNZ}

Finally, we wish to split our three candidate selections into two catalogues, a  $z<2.2$ `tracer sample' and a $z>2.2$ LyA sample using photometric redshifts for use by the 4MOST Cosmology Redshift Survey. These photometric redshifts will also be useful for projects to be discussed in Paper II. To determine the photometric redshifts  we utilize the ANNz2 software \citep{SAL2016}. This code uses artificial neural networks and boosted decision/regression trees to optimize the photo-z estimation and has already been implemented as part of the analysis in DES \citep{DES2014}. ANNz2 utilizes training based machine learning methods to derive the relationship between photometric observables and redshift.

\subsection{ANNz2 Training}
\label{sec:ANNz2training}

\begin{figure}
	\centering
	\includegraphics[width=\columnwidth]{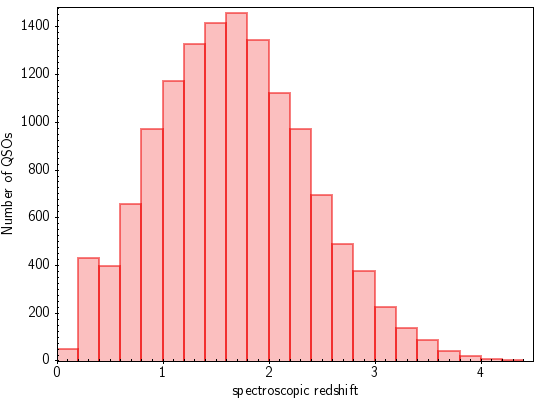} 
	\caption[Spectroscopic redshift of the training sample]{Spectroscopic redshift distribution of the full ANNz2 training and evaluation sample, using the DESI Guadalupe QSO Catalogue which will be released in DR1.}
	\label{fig:training_specz.png}
\end{figure}

To use ANNz2, we must train the algorithm with existing data which has similar properties to our candidates. We generate a training catalogue with the DESI Guadalupe data  over the $\approx$144 deg$^2$ overlap area discussed in Section \ref{sec:DESI_comparison}. From this, we use the 13374 QSOs in the overlapping area with the ATLAS NGC area, matched to the full VST-ATLAS data in order to train on photometry which we will be using for our dataset. The spectroscopic redshift distribution of the sample is shown in Fig.~\ref{fig:training_specz.png}. We use the ATLAS+`unWISE (neo6)' $ugrizW1W2$ magnitudes, errors, and the DESI spectroscopic redshifts to train the algorithm as these spectroscopically confirmed quasars were targeted through similar colour selections and cover the required redshift range.

\begin{figure}
	\centering
	\includegraphics[width=\columnwidth]{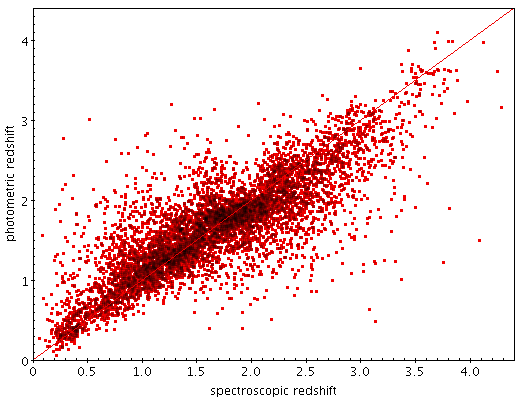} 
	\caption[Photo-z vs. spec-z of the training sample]{Photometric redshift compared to spectroscopic redshift for a random half of our DESI DR1 QSO sample used to test the training of the ANNz2 algorithm.}
	\label{fig:training_photoz_specz.png}
\end{figure}

To test the efficiency of the algorithm as well as our training sample, we divide the sample randomly in half, training on one half and testing the code on the other half. The result of that testing is seen in Fig.~\ref{fig:training_photoz_specz.png}. Here we plot the photometric redshift estimated by ANNz2 vs. the spectroscopic redshift of the testing half of the training sample.  We can compare our results with Fig. 4 of \cite{Yang2017}, which shows their photo-z vs. spec-z results using optical only as well as optical $+$ mid-infrared photometry. The top right panel of their Fig. 4 uses SDSS $ugriz$ and AllWISE W1, W2 to generate photometric redshifts, similar to our $ugriz$ and neo6 W1 and W2 data. Comparing our Fig.~\ref{fig:training_photoz_specz.png} to this top right panel in Figure 4 of \cite{Yang2017}, we see a similar relative degeneracy in the $0.8<z<3$ range. However, our version, which  has the benefit of deeper W1 and W2 data, seems to  represent an improvement due to the removal of the outlying clumps of photo-z degeneracies. 

We estimate the photometric redshift error by  measuring the standard deviation of $\Delta z= (z_{\rm photo}-z_{\rm spec})$ to be $\sigma_z=0.4$. We also find the standard deviation on the quantity $\frac{\Delta z}{1+z_{\rm spec}}$ to be $\pm0.16$. At $0.5<z_{\rm photo}<2.2$ this error is minimised and is reasonably constant at $\sigma_z\approx0.33$. 

\subsection{Photometric redshift samples}
\label{sec:photo_samp}

\begin{figure}
	\centering
	\includegraphics[width=\columnwidth]{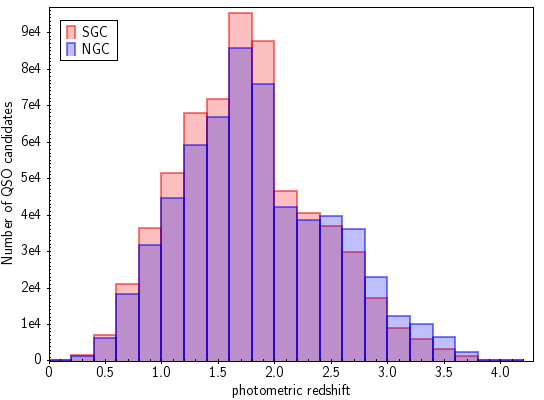} 
	\caption[The photo-z distribution of the final QSO catalogue]{Photo-z distribution of our total QSO candidate catalogue over the full VST-ATLAS footprint, split into NGC and SGC sub-samples.}
	\label{fig:Full_QSO_photoz.pdf}
\end{figure}

Fig.~\ref{fig:Full_QSO_photoz.pdf} shows the resulting ANNZ2 photometric redshift distributions of all ATLAS quasar candidates in the NGC and SGC. With ANNz2, we are also able to create $z_{\rm photo}<2.2$ and $z_{\rm photo}>2.2$ quasar candidate targets. The candidate sky densities for both samples are shown in Table~\ref{tab:ATLAS_ANNz_numbercounts}, where they are further split into NGC and SGC sky densities. We see that the overall $z_{\rm photo}<2.2$ sky  density is 193.5 deg$^{-2}$ with the NGC now being 18\% larger than the SGC (212.2 vs. 179.5 deg$^{-2}$), similar to the Priority 1 case. The $z_{\rm photo}>2.2$ sky density is 65.8 deg$^{-2}$ with the NGC now being 55\% higher than the SGC (82.3 vs. 53.1 deg$^{-2}$).  In Fig.~\ref{fig:Full_QSO_gband} (c) we show the number-magnitude relations for these two redshift ranges with the NGC-SGC-model comparison for $z_{\rm photo}<2.2$ being similar to the results previously found in Figs.~\ref{fig:Full_QSO_gband} (a,b). The 55\% higher NGC sky density for $z_{\rm photo}>2.2$ is due to artefacts in the non-UVX selection (as we can see in Table~\ref{tab:ATLAS_final_numbercounts}) and is seen over a wide magnitude range ($19<g<22.5$). 

From the candidate sky densities for these $z_{\rm photo}<2.2$ and $z_{\rm photo}>2.2$ catalogues,  we can estimate their true QSO sky densities. From  Table \ref{tab:ngcf1f2a}, our ATLAS efficiency decreases to 53\% at $z<2.2$ and 59\% at $z>2.2$ when we correct for ATLAS completeness using DESI, and our AAT 2dF observations suggest an efficiency of $\approx50$\%. Then, since DESI  completeness corrected ATLAS contaminations are 47\% at $z<2.2$ and 41\% at $z<2.2$, averaging gives efficiencies of 51.5\% at $z<2.2$ and 54.5\% at $z>2.2$, implying true sky densities of 100 deg$^{-2}$ at $z<2.2$ and 36 deg$^{-2}$ at $z>2.2$. So assuming at $z<2.2$ a 51.5\% efficiency and a candidate density from Table 9 of 193.5 deg$^{-2}$ also gives a QSO density of $\approx100$ deg$^{-2}$ (coincidentally). 

Adding eROSITA X-ray data then will give an increased stellar+extended sky density of $\approx45$\% over DESI and $\approx23$\% over a nominal $grW$ cut. Assuming an average increase of $\approx33$\% will then raise our $z<2.2$ sky density to $\approx130$ deg$^{-2}$. Our $z>2.2$ sky density is unaffected by the X-ray data and so will remain at $\approx36$ deg$^{-2}$. 

\begin{figure}
	\centering
	\includegraphics[width=\columnwidth]{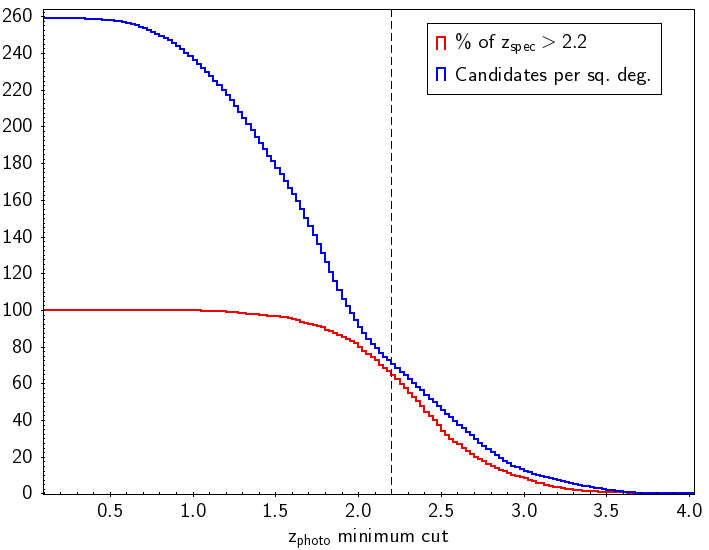}
	\caption[The photo-z distribution of the final QSO catalogue]{Dependence of QSO fractional completeness (as a percentage) with respect to the 1733 $z>2.2$ QSOs detected by ATLAS (red line) and sky density of candidates per deg$^{-2}$ (blue line) on $z_{\rm photo}$ minimum cut. Dashed line marks the nominal $z_{\rm photo}>2.2$ cut. }
	\label{fig:zphoto_cut}
\end{figure}

Now it should be noted that these estimates are approximate  because they do not take into account inaccuracies in the photometric redshifts. This is particularly true for the $z>2.2$ sample as can be seen in Fig. \ref{fig:zphoto_cut} where the fractional completeness with respect to the 1733 DESI QSOs detected by ATLAS (see Table \ref{tab:ATLAS_DESI_u_W2}) and candidate sky density are shown as a function of the $z_{\rm photo}$ cut. We find that the best trade-off between these two to is with a cut at $z_{\rm photo}>1.9$ where the $z>2.2$ fractional completeness is 90\% and the sky density is $\approx100$ deg$^{-2}$. We find that this adjustment of the $z_{\rm photo}$ cut is less of a consideration when the aim is to target a $z<2.2$ sample. Otherwise, we note that the highest overall completeness allied to the lowest overall candidate sky density at $z>2.2$ will always be achieved by making a combined redshift survey of the two photo-z samples simultaneously, because this avoids target duplication in the $1.9<z_{\rm photo}<2.2$ range.
 
\begin{figure*}
	\centering
\includegraphics[width=\textwidth]{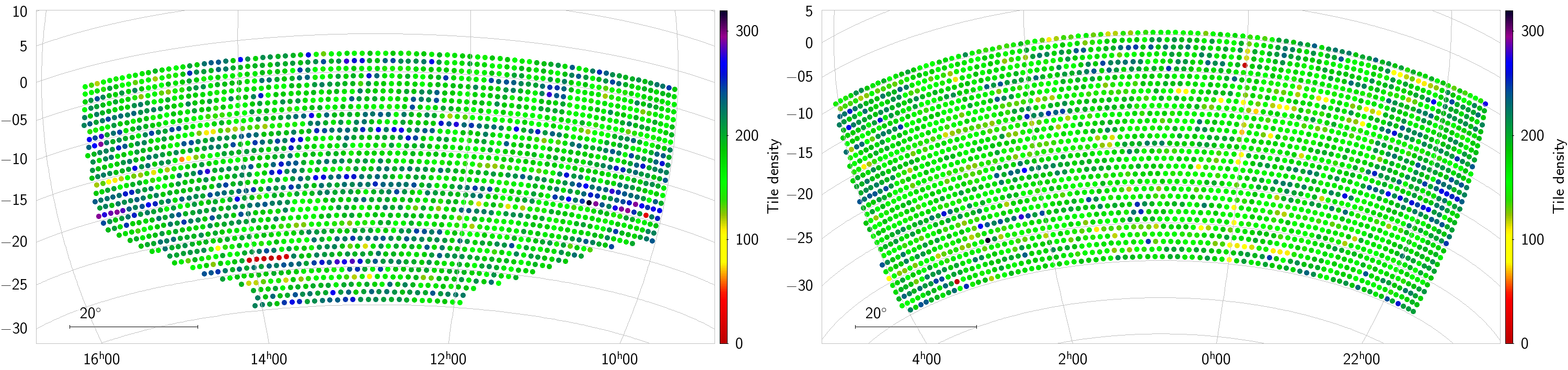}
	\caption[Tile density of total QSO candidates in the NGC]{ (a) VST-ATLAS tile density (deg$^{-2}$) for $z_{\rm photo}<2.2$ QSO candidates in the NGC. (b) VST-ATLAS tile density (deg$^{-2}$) for $z_{\rm photo}<2.2$ QSO candidates in the SGC.}
	\label{fig:NGC_SGC_qso_photoz_lo}
\end{figure*}

\begin{figure*}
	\centering
\includegraphics[width=\textwidth]{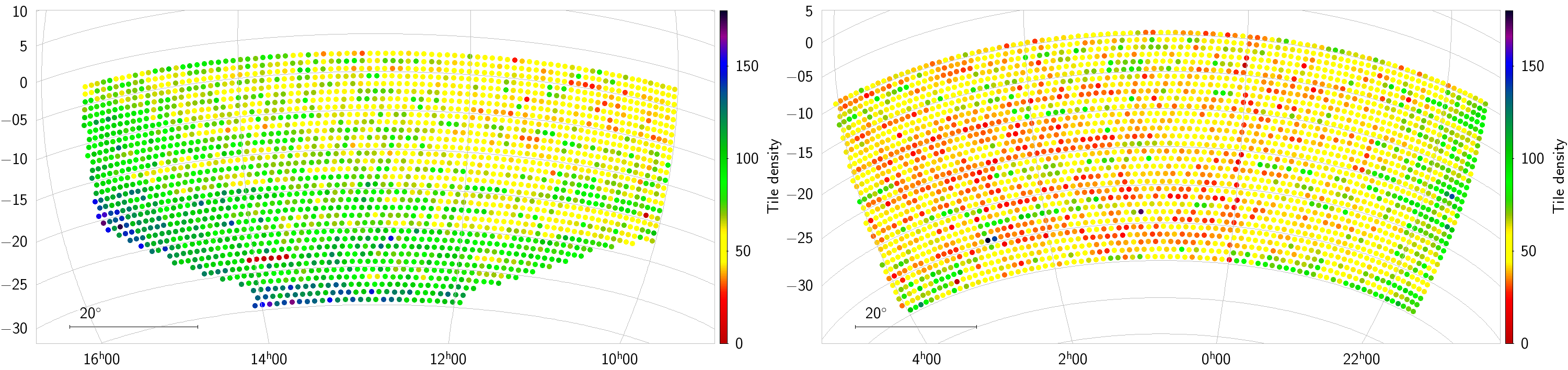}
	\caption[Tile density of total QSO candidates in the SGC]{(a) VST-ATLAS tile density (deg$^{-2}$) for $z_{\rm photo}>2.2$ QSO candidates in the NGC. (b) VST-ATLAS tile density (deg$^{-2}$) for $z_{\rm photo}>2.2$ QSO candidates in the SGC.}
	\label{fig:NGC_SGC_qso_photoz_hi}
\end{figure*}

Finally, in Figs.~\ref{fig:NGC_SGC_qso_photoz_lo} (a,b) we show the tile density maps for the $z<2.2$ sample in the NGC and SGC and in Figs.~\ref{fig:NGC_SGC_qso_photoz_hi} (a,b) we similarly show the tile density maps for the $z>2.2$ sample. While the $z<2.2$ maps look reasonably uniform across the sky as does the $z>2.2$ map in the SGC, the $z>2.2$ NGC map shows evidence of a gradient indicating that the high sky densities seen in Table~\ref{tab:ATLAS_final_numbercounts} and Fig.~\ref{fig:Full_QSO_gband} (c) are coming from the NGC at lower galactic latitudes. This could be due to extra star contamination, despite the fact that the main QSO contaminant is expected to be compact galaxies. Otherwise, the gradient might be due to some inaccuracy in our dust extinction correction.

\begin{table}
    \caption{QSO number counts and sky densities from our three selections (Priority 1, star $grW$ non-UVX, extended)  applied to the full VST-ATLAS footprint, divided into $z_{\rm photo}<2.2$ and $z_{\rm photo}>2.2$ candidates based on ANNz2.}
    \begin{tabular}{ |p{2.5cm}|p{2cm}|p{2cm}| }
      \hline
      Sky Area & total candidates & total candidates \\
               & $z_{\rm photo}<2.2$ & $z_{\rm photo}>2.2$ \\
      \hline
      NGC (2034 deg$^2$) & 431587 & 168151 \\
      NGC (deg$^{-2}$) & 212.2 deg$^{-2}$ & 82.3 deg$^{-2}$ \\
      SGC (2706 deg$^2$) & 485670 & 143580 \\
      SGC (deg$^{-2}$) & 179.5 deg$^{-2}$ & 53.1 deg$^{-2}$ \\
      total sky (4740 deg$^2$) & 917257 & 311731\\
      total sky (deg$^{-2}$) & 193.5 deg$^{-2}$ & 65.8 deg$^{-2}$ \\
     \hline
    \end{tabular}
    \label{tab:ATLAS_ANNz_numbercounts}
\end{table}

\section{Conclusions}
\label{sec:conclusions}

The main  aim of this paper was to present the VST-ATLAS QSO catalogue. We initially followed the photometric QSO selection work of \cite{Chehade2016} who used early VST ATLAS $ugriz$ data, combining it with AllWISE W1 and W2 survey  data.  These datasets differ from those of \cite{Chehade2016} in that the  sky coverage of ATLAS is now complete over its $\approx$4700 deg$^2$ and also  in the depth of the $u$-band which generally  has  240s exposure, $2\times$ more than \cite{Chehade2016} mainly by virtue of the ATLAS Chilean Survey (ACE, Barrientos et al in prep.). In addition to the VST ATLAS $ugriz$ photometry we have also replaced AllWISE with the WISE 6 year neo6 $W1$ and $W2$ MIR data \citep{Meisner2021} that has $6\times$ the 1 year exposure time of AllWISE. The neo6 W1 and W2 bands thus reach $\approx1$ mag fainter than the AllWISE survey data used by  \cite{Chehade2016}. We have also almost completed the DES $u$ Chile Extension (DEUCE) which provides $u$-band coverage to similar depth over a further 2800 deg$^2$ of the DES survey, which will allow full $ugrizW1W2$ photometry over the full $\approx$7500 deg$^2$ of the 4MOST Cosmology Redshift Survey (4CRS).

Here, we first used higher signal-noise William Herschel Deep Field (WHDF) $ugri$ and SpiES W1, W2 data \citep{Timlin2016} to establish the potential QSO sky density available to $g=22.5$. WHDF has the benefit of a 75ksec Chandra exposure to help us assess what the eROSITA X-ray data might add to the 4CRS survey, in terms of further increasing the QSO sky density. We also determined that the inclusion of objects that have been morphologically classed  as  extended sources and with QSO $ugriW1W2$ colours, provided the most complete QSO catalogue. Some of these sources are confirmed as extended sources even at HST $0.''1$ resolution due to host galaxy contributions and gravitational lensing. Overall, WHDF data  suggested that to $g<22.5$ a QSO sky density of $\approx269\pm67$ deg$^{-2}$ was in principle available when X-ray and optically extended sources are included. 

\subsection{QSO statistics in the full ATLAS catalogue}
\label{sec:full}

Armed with the lessons learned from these analyses, we applied our selections to VST ATLAS $ugri$  data, complemented by the W1 and W2 bands from the unWISE neo6 data release. We then suitably adjusted the cuts for ATLAS's less accurate photometry. This resulted in the full  VST ATLAS QSO catalogue containing $\approx1.2$ million QSO candidates with a sky density of $\approx$259 deg$^{-2}$.

Despite the WHDF results suggesting that $grW$ selections were the most complete, for VST ATLAS we still found improved efficiency and completeness when $u$ band selections were included. The reason  is that at VST ATLAS depth, despite ATLAS's excellent sub-arcsecond seeing in $g$ and $r$, star-galaxy separation gets increasingly unreliable as we approach our $g=22.5$ limit. Although the $grW$ selection  removes the main Galactic star populations with high efficiency, QSOs occupy the same $grW1$ locus as late-type galaxies and the more compact of these (NELGs) comprise our main contamination. We note that a $W1-W2>0.4$ criterion can also reduce galaxy contamination but the neo6 W2 data runs out $\approx0.7$ mag before our $g=22.5$ limit.  DESI take advantage of their deeper `forced' W1 and W2 data to eliminate more galaxies. Here, we instead  exploit our relatively deep `forced'  $u$ data. We demand our candidates  pass  our joint UVX $+$ $grW$ cuts and similarly our joint $grW$ $+$ non-UVX  cuts. This reduces our selected stellar QSO candidate  density by $\approx60$\%. If, instead of including the $u$ data, we apply the strict $W1-W2>0.4$ cut to the $grW$ selection, effectively now demanding a $W2$ detection for all candidates, the sky density shows a bigger reduction to 147 deg$^{-2}$. However, in the latter case the completeness compared to DESI also drops from 83\% to 78\% including a 17\% drop in the $z>2.2$ range. So $u$ data appears to most advantage the $z>2.2$ sample in the VST ATLAS survey.

We then used  spectroscopically confirmed 2QZ, 2QDESp, SDSS eBOSS, DESI QSOs and also new, specially commissioned, 2dF observations of ATLAS QSO candidates to test our selections. The latter 2dF results suggest that we shall reach at least $>110$ deg$^{-2}$ at $z<2.2$ and $>30$ deg$^{-2}$ at $z>2.2$. But note these results assume we apply UVX + $grW$ and non-UVX cuts simultaneously; if only the non-UVX cut was made a significantly lower $z>2.2$ density would be found.

We find a completeness of 70\% with respect to confirmed DESI QSOs for our total candidate sample, with an efficiency of 48\%. Through comparing with 2QZ, 2QDESp, and eBOSS, we are able to see good completeness of $\approx88\%$, with the brightest, stellar, ATLAS selections giving $\approx97\%$ completeness.

We performed a $g-$band number count comparison with the work done by \cite{PD2016}. These models are also used by DESI to determine their expected QSO number counts. We find that our observed number counts of 259 deg$^{-2}$ at $g<22.5$ are somewhat higher than the QSO sky density of 195 deg$^{-2}$ predicted by their PLE$+$LEDE QSO luminosity function  model. However, when the estimated efficiency and  incompleteness of our sample is taken into account, the ATLAS QSO counts are expected to be in reasonable agreement with the model, although still lower than the WHDF QSO sky density of $\approx269\pm67$ deg$^{-2}$, given the WHDF's advantage of having much deeper Chandra X-ray, Spitzer SpiES W1 and W2 and optical data available.

\subsection{QSO statistics in ATLAS catalogues split at $z_{\rm photo}=2.2$ }
\label{sec:photo}
Applying the ANNZ2 algorithm of  \cite{SAL2016} to our final QSO candidate catalogue provided photometric redshift estimates for all catalogue members. The resulting QSO candidate sky density over our full $\approx$4740 deg$^2$ is 194 deg$^{-2}$ for the $z<2.2$ `tracer' QSO candidate catalogue and 66 deg$^{-2}$ for the $z>2.2$ LyA QSO candidate catalogue.

We then estimated the true QSO sky densities for the ATLAS catalogues split at $z_{\rm photo}<2.2$ and $z_{\rm photo}>2.2$, finding true sky densities of 100 deg$^{-2}$ at $z<2.2$ and 36 deg$^{-2}$ at $z>2.2$. Adding eROSITA X-ray data will then increase our $z<2.2$ sky density to $\approx130$ deg$^{-2}$ with our $z_{\rm photo}>2.2$ sky density remaining at $\approx36$ deg$^{-2}$. 
These estimates ignore the $\pm0.4$ photo-$z$ error and from Fig. \ref{fig:zphoto_cut} we found that the best trade-off between completeness and efficiency in our high redshift sample is with a cut at $z_{\rm photo}>1.9$. Otherwise, we note that the highest overall completeness coupled with the lowest overall candidate sky density at $z>2.2$ is best achieved via a combined redshift survey of the two photo-$z$ samples simultaneously where there is no need to incur duplication of targets e.g. in the $1.9<z_{\rm photo}<2.2$ range.

\subsection{Future applications of the VST ATLAS QSO catalogues}
\label{sec:future}

Further improvements to VST ATLAS QSO selection including deeper  NEOWISE data and also upcoming eROSITA X-ray data, mean that we are well positioned to exceed our target QSO sky densities of 130 deg$^{-2}$ at $z<2.2$ and 30 deg$^{-2}$ at $z>2.2$. Although the ATLAS QSO catalogues already include photometric redshifts that are accurate to $\sigma_z=0.4$, more accurate spectroscopic redshifts will be needed to measure Redshift Space Distortions (RSD) and Baryon Acoustic Oscillation (BAO) scales from QSO and Lyman $\alpha$ forest clustering to make the most accurate measurements of cosmological parameters. 

This ATLAS QSO catalogue will therefore ultimately be used as a basis for the  QSO component of the 4MOST Cosmology  Redshift Survey.  With the addition of the DES area, this 4CRS QSO redshift  survey will cover 7500 deg$^2$ of sky with a QSO target sky density of 240deg$^{-2}$.  The 4MOST eROSITA AGN survey will also cover most of this area and contribute $\approx$55 deg$^{-2}$ or $\approx$40\% of the target $z<2.2$ `tracer' QSO sky density of $\approx$130 deg$^{-2}$. Thus by combining the ATLAS optical/MIR and eROSITA X-ray QSO surveys, we can produce a QSO redshift survey that is highly competitive for cosmology at a much reduced cost. As well as providing high-quality BAO and RSD  measurements out to $z\approx3.5$, the 4MOST QSO redshift survey will also  give vital support to DES and LSST galaxy weak lensing analyses at lower redshift ($z\la1$) by constraining  the crucial redshift distribution of the lensed galaxies via QSO-galaxy cross-clustering. 

Meanwhile, in advance of 4CRS, in Paper II (Eltvedt et al, in preparation) we shall exploit the  current ATLAS QSO photo-z catalogue to measure QSO lensing magnification by foreground galaxies as well as galaxy clusters from the VST ATLAS Galaxy Cluster Catalogue I (\citealt{Ansarinejad2022}) using the cross-correlation technique. We shall also similarly report on measuring the magnification of Cosmic Microwave Background fluctuations lensed by the QSOs themselves and combine all of these results to make new  estimates of cosmological parameters.

\section*{Acknowledgements}

We acknowledge use of the ESO VLT Survey Telescope (VST) ATLAS. The ATLAS survey is based on data products from observations made with ESO Telescopes at the La Silla Paranal Observatory under program ID 177.A-3011(A,B,C,D,E.F,G,H,I,J,K,L,M,N) (see \citealt{Shanks2015}).

We acknowledge the use of  data products from WISE, which is a joint project of the University of California, Los Angeles, and the Jet Propulsion Laboratory (JPL)/California Institute of Technology (Caltech), funded by the National Aeronautics and Space Administration (NASA), and from NEOWISE, which is a JPL/Caltech project funded by NASA. 

We acknowledge the use of  SPIES survey observations made with the Spitzer Space Telescope, which is operated by the Jet Propulsion Laboratory, California Institute of Technology under a contract with NASA. 

We acknowledge use of SDSS imaging and spectroscopic data. Funding for SDSS-III has been provided by the Alfred P. Sloan Foundation, the Participating Institutions, the National Science Foundation and the US Department of Energy Office of Science. 

We further acknowledge use of the DESI Guadalupe spectroscopic data which will be released in DR1. DESI is supported by the Director, Office of Science, Office of High Energy Physics of the U.S. Department of Energy under Contract No. DE–AC02–05CH11231, and by the National Energy Research Scientific Computing Center, a DOE Office of Science User Facility under the same contract; additional support for DESI is provided by the U.S. National Science Foundation, Division of Astronomical Sciences under Contract No. AST-0950945 to the NSF’s National Optical-Infrared Astronomy Research Laboratory; the Science and Technologies Facilities Council of the United Kingdom; the Gordon and Betty Moore Foundation; the Heising-Simons Foundation; the French Alternative Energies and Atomic Energy Commission (CEA); the National Council of Science and Technology of Mexico; the Ministry of Economy of Spain, and by the DESI Member Institutions.

We thank all staff at the Anglo-Australian Telescope (AAT) for their assistance with observations using the 2dF+AAOmega instruments. We thank OPTICON and ATAC for their financial support of our AAT observations.

BA acknowledges support from the Australian Research Council’s Discovery Projects scheme (DP200101068).

We finally acknowledge STFC Consolidated Grant  ST/T000244/1 in supporting this research.

For the purpose of open access, the author has applied a Creative Commons Attribution (CC BY) licence to any Author Accepted Manuscript version arising.

\section*{Data Access Statement}


The ESO VST ATLAS, WISE and Spitzer SPIES   data we have used are all
publicly available. In the case of the DESI data used here, this will be made publicly available via the regular public data releases scheduled by the DESI collaboration. All other data relevant to this publication will be supplied on request to the authors. 



\bibliographystyle{mnras}
\bibliography{bibliography_v3_TS_newest} 


\appendix

\section{WHDF X-ray and DESI QSOs}
\label{appendix}
Here we list the QSO contents of the WHDF from the Chandra X-ray source list of \cite{Bielby2012} 
in Table \ref{tab:WHDF_bielby_fullinfo} and from preliminary DESI QSO redshift survey data in
Table \ref{tab:WHDF_desi_fullinfo}.
\begin{table*}
    {\centering}
    \caption[Photometric, morphological and redshift information for the $15$  X-ray QSOs from \protect\cite{Bielby2012}.]{Full colour, morphology and redshift information for the 15 X-ray QSOs from \protect\cite{Bielby2012} found in the WHDF. X-ray absorbed QSOs  are   bolded. (D) in second column indicates also detected by DESI (see Table \ref{tab:WHDF_desi_fullinfo}). The $ugriz$ magnitudes come from the SDSS Stripe 82 data, the W1 and W2 fluxes are from \cite{Timlin2016}, and the X-ray fluxes are from \cite{Vallbe2004}.}
    \begin{tabular}{ p{2.0cm}||p{1.5cm}|p{0.7cm}|p{0.7cm}|p{0.7cm}|p{0.7cm}|p{0.7cm}|p{0.7cm}|p{0.7cm}|p{1.5cm}|p{1.0cm}}
      \hline
       ID & morphology & u & g & r & i & z & W1 & W2 & $S_X(0.5-10)$&Redshift\\
      \hline
WHDFCH005          & star      & 21.19 & 20.83 & 20.86 & 20.61 & 20.56 & 16.47 & 15.58& $5.62\times10^{-14}$ & 0.52 \\
\textbf{WHDFCH007} & galaxy    & 23.73 & 23.22 & 22.88 & 22.27 & 21.87 & 15.98 & 15.14& $1.17\times10^{-14}$ & 1.33 \\
\textbf{WHDFCH008} & galaxy    & 23.84 & 24.00 & 24.01 & 23.38 & 22.32 & 17.14 & 16.42& $3.62\times10^{-15}$ & 2.12 \\
WHDFCH016          & star (D)  & 20.95 & 20.67 & 20.61 & 20.34 & 20.29 & 17.20 & 16.15& $1.44\times10^{-14}$ & 1.73 \\
WHDFCH017          & star (D)  & 20.24 & 20.04 & 19.85 & 19.60 & 19.11 & 15.12 & 14.54& $3.22\times10^{-13}$ & 0.40 \\
WHDFCH020          & galaxy    & 22.34 & 22.05 & 21.67 & 21.35 & 20.86 & 16.50 & 16.15& $1.09\times10^{-14}$ & 0.95 \\
WHDFCH036          & star (D)  & 22.14 & 21.62 & 21.68 & 21.46 & 21.00 & 16.18 & 15.56& $6.26\times10^{-14}$ & 0.83 \\
\textbf{WHDFCH044} & star (D)  & 22.73 & 21.84 & 20.49 & 19.83 & 19.14 & 13.54 & 12.55& $2.66\times10^{-14}$ & 0.79 \\
WHDFCH048          & galaxy (D)& 23.16 & 22.57 & 22.13 & 21.67 & 21.76 & 16.30 & 15.42& $2.15\times10^{-14}$ & 1.52 \\
WHDFCH055          & star      & 23.65 & 22.28 & 21.73 & 21.15 & 20.60 & 16.26 & 15.97& $2.17\times10^{-14}$ & 0.74 \\
WHDFCH090          & star (D)  & 21.07 & 21.03 & 20.62 & 20.72 & 20.76 & 16.26 & 15.47& $4.83\times10^{-14}$ & 1.32 \\
\textbf{WHDFCH099} & star (D)  & 20.52 & 20.34 & 20.25 & 20.23 & 20.00 & 15.58 & 14.96& $8.84\times10^{-15}$ & 0.82 \\
WHDFCH109          & star      & 18.39 & 18.07 & 18.14 & 18.00 & 18.13 & 13.80 & 12.95& $6.69\times10^{-14}$ & 0.57 \\
WHDFCH110          & galaxy    & 22.73 & 21.91 & 21.22 & 20.59 & 19.95 & 15.50 & 15.42& $2.20\times10^{-14}$ & 0.82 \\
WHDFCH113          & star (D)  & 22.19 & 21.56 & 21.51 & 21.59 & 21.44 & 18.30 & 17.47& $5.99\times10^{-15}$ & 2.54 \\

     \hline
    \end{tabular}
    \label{tab:WHDF_bielby_fullinfo}
\end{table*}

\begin{table*}{\centering}
    \caption[Colour, morphology and redshift information for the QSOs found in the WHDF by SDSS and DESI.]{Photometric, morphological and redshift information for the 13 QSOs from DESI in the WHDF. In the first column, bracketed names are those for DESI sources detected in Chandra X-rays at $>3\sigma$ by \protect\cite{Vallbe2004} but not listed by \protect\cite{Bielby2012}. In second column, (X) indicates also listed as an  X-ray QSO by  \protect\cite{Bielby2012} (see Table \ref{tab:WHDF_bielby_fullinfo}); Column 10: (-) indicates no X-ray detection at 3$\sigma$. All other fluxes not listed by \protect\cite{Bielby2012}  are from \protect\cite{Vallbe2004}.  We note that the DESI data used here is preliminary and subject to change in final, public,  DESI data releases.}
    \begin{tabular}{ p{2.5cm}||p{2.0cm}|p{0.7cm}|p{0.7cm}|p{0.7cm}|p{0.7cm}|p{0.7cm}|p{0.7cm}|p{0.7cm}|p{1.5cm}|p{1.0cm}}
      \hline
    WHDF ID & morphology & u & g & r & i & z & W1 & W2 & $S_X(0.5-10)$ & Redshift\\
      \hline
1109              & star      & 24.94 & 22.45 & 22.23 & 22.06 & 22.14 & 17.84 & 17.27 & - & 3.087 \\
3630              & star (X)  & 21.07 & 21.03 & 20.62 & 20.72 & 20.76 & 16.26 & 15.47 &  $4.83\times10^{-14}$ & 1.334 \\
2779  (WHDFCH038) & star      & 23.47 & 21.26 & 21.05 & 21.11 & 21.24 & 19.00 & 17.68 & $6.30\times10^{-15}$ & 3.138 \\ 
8222  (WHDFCH014) & galaxy    & 23.22 & 22.34 & 21.96 & 21.82 & 21.46 & 17.24 & 16.68 & $7.10\times10^{-15}$ & 2.679 \\
254               & star      & 24.47 & 22.70 & 21.83 & 21.69 & 21.48 & 17.68 & 16.36 & - & 2.593 \\
5964              & star (X)  & 20.24 & 20.04 & 19.85 & 19.60 & 19.11 & 15.12 & 14.54 & $ 3.22\times10^{-13}$  & 0.397 \\
10665             & star (X)  & 22.73 & 21.84 & 20.49 & 19.83 & 19.14 & 13.54 & 12.55 & $2.66\times10^{-14}$ & 0.799 \\
8779              & star (X)  & 22.19 & 21.56 & 21.51 & 21.59 & 21.44 & 18.30 & 17.47 & $5.99\times10^{-15}$ & 2.544 \\
14697             & galaxy (X)& 23.16 & 22.57 & 22.13 & 21.67 & 21.76 & 16.30 & 15.42 & $2.15\times10^{-14}$ & 1.539 \\
14428             & star (X)  & 22.14 & 21.62 & 21.68 & 21.46 & 21.00 & 16.18 & 15.56 & $6.26\times10^{-14}$ & 0.833 \\
11642             & star (X)  & 20.52 & 20.34 & 20.25 & 20.23 & 20.00 & 15.58 & 14.96 & $8.84\times10^{-15}$ & 0.820 \\
5971              & star (X)  & 20.95 & 20.67 & 20.61 & 20.34 & 20.29 & 17.20 & 16.15 & $1.44\times10^{-14}$  & 1.753 \\
3081  (WHDFCH052) & galaxy    & 21.77 & 21.73 & 21.56 & 21.62 & 21.66 & 17.26 & 16.59 & $1.20\times10^{-14}$ & 1.306 \\
     \hline
    \end{tabular}
    \label{tab:WHDF_desi_fullinfo}
\end{table*}






\bsp	
\label{lastpage}
\end{document}